# Cable Stability

*L. Bottura*[1]
CERN, Geneva, Switzerland

**Abstract**
Superconductor stability is at the core of the design of any successful cable and magnet application. This chapter reviews the initial understanding of the stability mechanism, and reviews matters of importance for stability such as the nature and magnitude of the perturbation spectrum and the cooling mechanisms. Various stability strategies are studied, providing criteria that depend on the desired design and operating conditions.

## 1 Introduction

The first superconducting magnets built in the decade around 1960 out of the newly available superconducting materials (mainly Nb–Zr, $Nb_3Sn$, and Nb–Ti, in the form of tapes or large monofilamentary strands) had their first transition to the normal state – that is, a *quench* – long before reaching the expected critical current, largely disappointing the constructors. This happened in spite of the success in the development of pure superconducting materials. The situation has been illustrated by Chester [1] in an excellent review article on the status of the development of superconducting magnets:

*"[...] the development of superconducting solenoids and magnets has been far from straightforward, mainly because the behaviour of the materials in coils frequently did not accord with the behaviour of short samples. [...] The large number of coils [...] wound from Nb–Zr and Nb–Ti wire, and [...] $Nb_3Sn$, revealed several intriguing and very frustrating characteristics of these materials in magnets."*

An example of this behaviour is shown in Fig. 1, reporting the history of the maximum current reached in a superconducting solenoid wound with Nb–Zr wire. At the first powering, the magnet reached only ~12 A, after which it quenched. At the following attempt to power the magnet, the current that could be reached before quench was higher. This process continued at each attempt, and the maximum current that could be reached increased quench after quench, slowly approaching a plateau – in the example of Fig. 1, at around 28 A. This behaviour became known as *training* [2], and the curve reported in Fig. 1 is called the *training curve*. The plateau current reached, however, was still below the expected maximum current-carrying limit of the cable. As shown in Fig. 2, coils wound from 0.25 mm Nb–Zr wire could achieve only a small fraction of the critical current of the single wire.

---

[1] luca.bottura@cern.ch

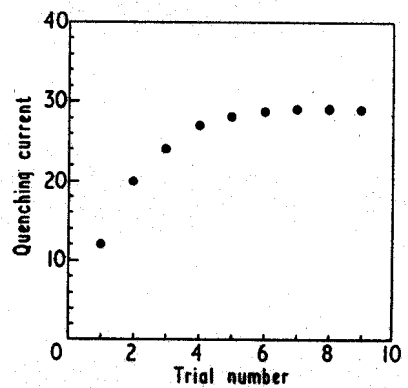

**Fig. 1:** The training curve for an early Nb–Zr superconducting solenoid. (Reproduced from [1].)

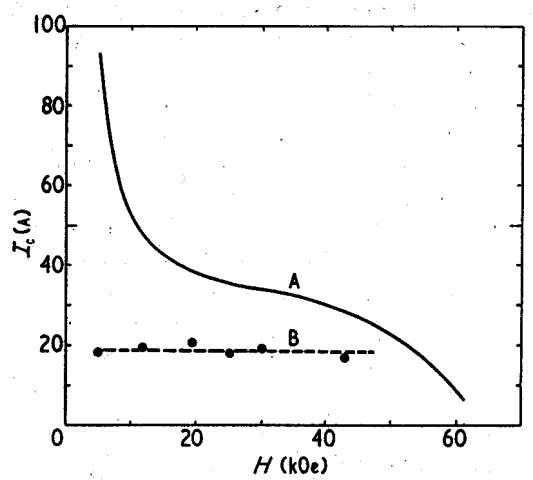

**Fig. 2:** The typical critical current of a Nb–Zr superconducting wire (curve A) compared with that in coils producing the same field (curve B): 10 kOe = 1 T. (Reproduced from [1].)

The current limitation observed was originally thought as originating from *bad spots* in the wires or cables, and thus attributed to bad homogeneity in the quality of the superconductor. This idea produced the concept of *degradation* of the performance of the conductor. Although training clearly showed that a physical degradation could not be responsible for the bad performance, the misleading name remained as an inheritance of the imprecise understanding. Particularly puzzling was the fact that the degradation depended on the coil construction and on its geometry. Again quoting Chester [1]:

*"The prediction of the degraded current for any new shape or size of coil proved to be impossible and, for a time, the development of coils passed through a very speculative and empirical phase."*

A principle not yet fully understood at the time was that of *stability* of the cable with respect to external disturbances. Insufficient stability and large external disturbances were the key issues in the failure of the early experiments on superconducting magnets. It has since become understood that a superconducting magnet is always subject to a series of energy inputs of very different natures, time-scales, and magnitudes, the so-called *disturbance spectrum* [3]. The energy input in the superconducting cable increases its temperature and can be sufficient to take the superconducting material above critical conditions, where it becomes resistive and Joule heating is generated. The region that has transited to the normal resistive state is the so-called *normal zone* in the magnet. Most materials at cryogenic temperature have a small heat capacity (ideally vanishing at absolute zero), and the temperature margin, the difference between the operating temperature $T_{op}$ and the temperature at which current sharing starts, $T_{cs}$, must be kept small to achieve an economic design. The result is that the energy necessary to produce a normal zone can be small, typically in the range of tens to hundreds of microjoules for a few cubic millimetres of strand.

If not prevented by other mechanisms, the temperature in the normal zone increases further and the normal front propagates, so that the superconductor cascades from its nominal operating point into an irreversible process, leading to a complete loss of superconductivity in the magnet; that is, the magnet experiences a quench. This sequence of events is shown schematically in Fig. 3.

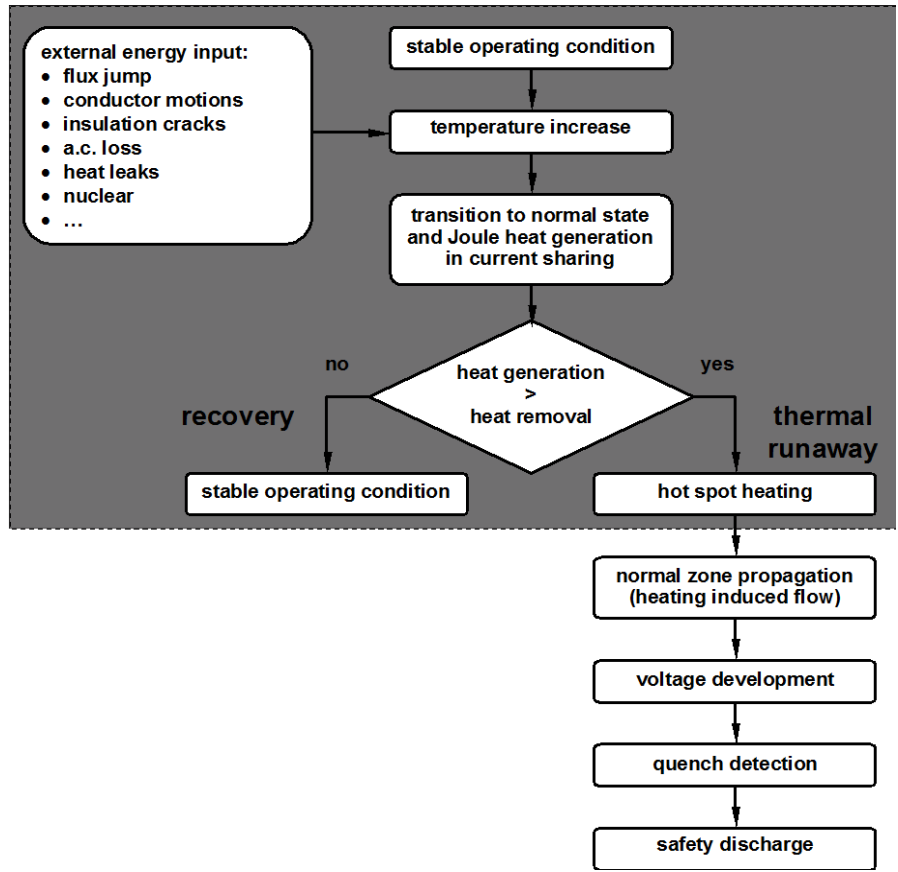

**Fig. 3:** An event tree following an external energy input, and leading from stable operating conditions back to stable operation or to a magnet quench. The stability design and analysis are concentrated on the shaded area in the event tree.

Even if properly protected against damage, a magnet quench is an undesirable event in terms of availability and cost. A well-designed magnet will not quench under normal operating conditions. The study of stability has evolved through many years of experimentation and analysis towards an understanding of the processes and mechanisms whereby a superconductor remains (or not) within its operating region, thus ensuring *stable* operation of the magnet; that is, free from quenches. With reference to the schematic representation of Fig. 3, stability is therefore mainly concerned with the phases in the event tree enclosed in the shaded area.

In spite of the substantial progress in understanding and improvement in the manufacturing techniques, stability is still one of the limiting factors for high-performance magnets. As an example, Fig. 4 reports the sequence of training quenches for an accelerator magnet, showing that, to achieve high performance, training may still be necessary, and still the short-sample limit is not reached. The training behaviour can differ significantly depending on manufacturing details such as the quality of the fitting of the mechanical structure and the coil windings.

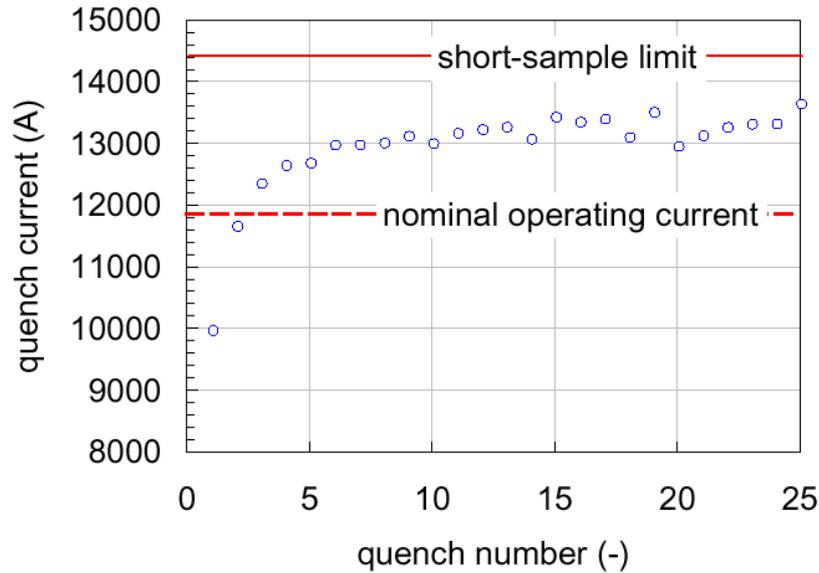

**Fig. 4:** A sequence of training quenches for a short model of an LHC dipole magnet, showing the initial, short training period necessary to reach nominal operating conditions and a plateau current reached after several quenches. Note that the plateau is below the expected short-sample limit of the cable.

This chapter deals with cable stability under most common conditions found in superconducting magnet design. The first step in a sound design is to estimate the envelope of the perturbations that will be experienced by the magnet. Typical energies and time-scales of the perturbation spectrum are reviewed in Section 3.2. The conductor can then be designed to accommodate these perturbations by means of a sufficiently large stability margin, using the concepts discussed in Sections 3.6 (adiabatic stabilization), 3.7 (cryostability), 3.8 (cold-end recovery and the equal-area theorem), 3.9 (well-cooled operation for force-flow cooled conductors), and 3.10 (minimum propagating zone). Appropriate examples based on existing magnets are given in each section. In order to develop the relevant concepts and techniques, we will introduce in Section 3.3 a general form of the heat balance for a superconducting cable, and in Sections 3.5 and 3.6 we will discuss the details of heat generation during current sharing and steady state, and transient heat transfer to a cryogenic coolant. Finally, in Section 3.11, we will discuss some advanced topics (current distribution) in stability design and analysis.

## 2   The perturbation spectrum

Several mechanisms can cause the generation of heat in a superconducting cable carrying a current in a magnetic field and operating in a cryogenic environment. These perturbations are distributed over a wide spectrum, with large differences in magnitude and time-scale depending on the origin of the perturbation itself. Some disturbances – for example, the mechanical energy release due to a small motion of a superconducting wire – can be extremely localized in space and time, and can initiate only normal zones of small volume. Other disturbances can affect large portions of a superconducting magnet and last a significant time – for example, a.c. loss for pulsed operation or heat deposition from nuclear processes in a superconducting accelerator magnet – and thus they can potentially produce normal zones with large volume. The specific perturbation spectrum for a particular application depends on the design of the superconducting magnet and on the operating conditions of the system, and is difficult to generalize.

In this section, we give a sample of most common disturbances, and their associated energy, volume and time-scales. Stability analysis and design is mostly concerned with transient perturbations of limited duration. Hence they can be characterized by the total energy deposited during the transient in units of [J]. For the comparison of different phenomena acting on different volumes, it is useful to reason in terms of energy density with reference to the total cross-section of the conductor. For practical reasons, this quantity is usually quoted in units of [mJ·cm$^{-3}$], as the typical values then range from fractions of the unit to a few hundreds.

As is the case for heat leaks from room temperature, continuous heat deposition is characterized by the heating power, measured in units of [W] for localized inputs, in units of [W·m$^{-2}$] for surface heat loads, or in units of [W·m$^{-3}$] for volumetric loads. These perturbations, although very important for the overall performance of a system, are usually not the concern for stability. They are dealt with by proper sizing of the thermal insulation and cooling system. It is nonetheless interesting to show the corresponding energy deposited over the typical time-scales of interest for stability, also discussed here in units of [mJ·cm$^{-3}$] to provide comparable dimensions.

## 2.1 Flux jumps

A small heat input into a superconductor in a magnetic field – due, for example, to any of the reasons discussed below – causes a decrease of the critical current density $J_c$ through the temperature increase (all technical superconductors have a negative $J_c(T)$ slope). In adiabatic conditions, the magnetization of the superconductor (proportional to the current density) also decreases, resulting in a penetration of the external magnetic field in the superconducting bulk. A part of the energy stored in the magnetization profile is therefore dissipated resistively within the superconductor. The energy release caused by the decrease of the magnetization can be sufficient to cause an irreversible transition of the wire to the normal state – a *flux jump*.

In order to estimate the maximum possible energy release during a flux jump, we can make the assumption that during the transient the magnetization of a superconducting filament disappears because of the process described above. To simplify matters in this study, we approximate the filament as a plane slab with thickness $D$ identical to the diameter of the filament. Following a ramp of increasing magnetic field up to a value $B$ much larger than the penetration field, and neglecting the influence of transport current, the shielding currents fill the slab, in the fully penetrated state.

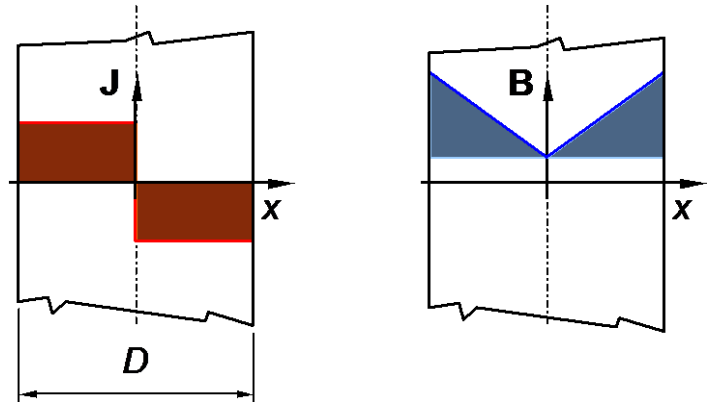

**Fig. 5:** The profile of the magnetic field and current density in a fully penetrated superconducting filament, ideally represented by a slab of thickness $D$ along $x$ and infinite extension in the other two dimensions. The area shaded in the plot of the field profile disappears during a complete flux jump.

The magnetic field contribution $\delta B$ due to the shielding currents has a dependence on the position $x$, given by

$$\delta B = \mu_0 J_c x , \qquad (1)$$

where $J_c$ is the critical current density of the superconductor. The field profile is shown in Fig. 5. Following a complete flux jump, the energy stored in this field profile is dissipated resistively inside the filament. The average energy dissipation $Q'''$ per unit volume of the filament can then be calculated as follows:

$$Q''' = \frac{2}{D} \int_0^{D/2} \frac{\delta B^2}{2\mu_0} \, dx = \frac{\mu_0 J_c^2 D^2}{24}. \tag{2}$$

From the above expression, we can estimate the energy density released during a flux jump in a typical superconductor such as Nb–Ti. Assuming that $J_c$ is 5000 A·mm$^{-2}$ and that the filament diameter $D$ is 50 μm, we obtain an energy density of 3 mJ·cm$^{-3}$. This energy can be released extremely rapidly, typically in tens of microseconds up to fractions of milliseconds, and is sufficient to increase adiabatically the temperature of a Nb–Ti strand by few degrees.

The magnetization of a fully penetrated superconducting filament is proportional to the size of the filament, and the energy stored in the trapped magnetic field is proportional to the square of the filament size, as shown by Eq. (2). Because of this, flux jumps were a principal problem at times when superconducting material technology did not allow the production of a fine subdivision of the superconductor in the wire. In fact, flux jumps were among the first perturbations recognized to be responsible for performance degradation, and thus were studied intensely in the late 1960s and early 1970s (see, e.g., the review by the Rutherford Laboratory Superconducting Applications Group [4]), leading to one of the first quantifications of the idea of a disturbance spectrum. Because of this, the early considerations of stability were often interleaved with flux jump theory.

Fine subdivision in small filaments is the most obvious cure for flux jumping. On the one hand, it reduces the energy that is dissipated, as is clear from Eq. (2), and therefore it eases the so-called *adiabatic* stabilization. On the other hand, fine subdivision means an increase of the superconductor surface, making it easier to remove the heat generated by the flux penetration in the bulk superconductor efficiently – so-called *dynamic* stabilization. Nowadays, flux jumps are no longer a problem for standard production of low-temperature superconducting wires (typically based on Nb–Ti and Nb$_3$Sn materials). Flux jumps may still play some role in high-temperature superconductors operated at low temperatures (around and below 20 K), although the steadily improving technology is quickly making this statement obsolete.

## 2.2 Mechanical events

A superconducting magnet is always subjected to stresses, from pre-loading at assembly, differential thermal contractions at cool-down, or resulting from the electromagnetic forces at operation. The forces acting on a superconducting cable can induce small movements. In some cases, the stress can be large compared to the elastic and failure limits of the materials, and displacement can take place as a result of material yield or fracture. Any displacement causes a change in the stress state associated with a release of a part of the mechanical energy stored. The release of the mechanical energy can happen locally, through micro-slips constrained by friction, material yielding, vibration, or local cracking [5].

We can appreciate the amount of mechanical energy associated with one of the above events by estimating the energy release due to a hypothetical strand motion, as shown schematically in Fig. 6. A strand operating at a current $I_{op}$ in a transverse field $B_{op}$ results in a Lorentz force $F'$ per unit length of the strand of

$$F' = I_{op} B_{op}. \tag{3}$$

Taking typical values for the bending magnets of a particle accelerator, an $I_{op}$ of 400 A and a $B_{op}$ of 10 T, the force per unit length is $F' = 4$ kN·m$^{-1}$. This force is reacted against the other wires in the winding pack and the structure of the magnet. However, even in a tightly packed winding, the wire can move through small distances because of the geometrical tolerances on the wire dimensions and the limitations on the control of the winding geometry. Movements $\delta$ of a few micrometres over a strand length $l$ of a few millimetres are not uncommon if the winding pack is not fully impregnated. The work $W$ performed by the Lorentz force during a movement $\delta$ can be calculated as follows:

$$W = F' \delta l. \tag{4}$$

A movement of 10 μm of a 1 mm length of a single strand under the conditions given above is associated with work $W = 40$ μJ. This work corresponds to an energy density of

$$W''' = \frac{F' \delta}{A_{strand}}, \tag{5}$$

where $A_{strand}$ is the total strand cross-section. For a strand with a 1 mm$^2$ cross-section, typical of the application considered, the energy density is 40 mJ·cm$^{-3}$. The mechanical work is partially dissipated as friction against the other wires and partially as a resistive loss induced by the electric field on the moving wire. The percentage of energy dissipation depends on the detail of the process and cannot be estimated easily, ranging from only a few per cent to large fractions of the above estimate. These events take place on a few cubic millimetres of cable, and are fast, with typical times that generally range from tens of microseconds to a few milliseconds.

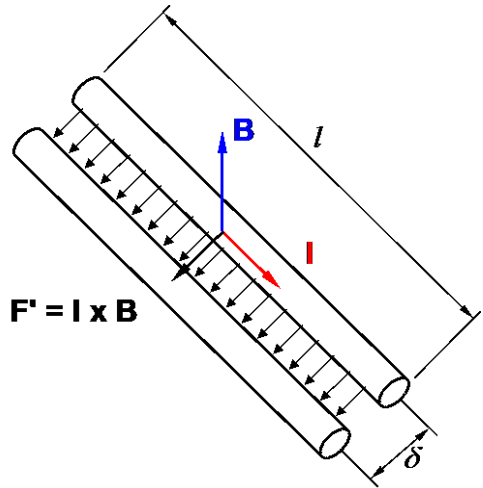

**Fig. 6:** A schematic representation of strand motion in a magnetic field. The strand length $l$ is subjected to a transverse electromagnetic load $F'$ per unit length and moves by a distance $\delta$ in the direction of the force.

In a few cases, the thermal or magnetic stresses become large enough to cause some part of the structural materials to yield, or to fracture, depending on the embrittlement at low temperature. Under these circumstances, more massive disturbances are produced, such as de-bonding or shear failure of the insulation or displacement of a part of the magnet. Heat is released in these processes through friction during motion, or once the movement is stopped. These *massive* processes are, however, rarely taking place at the single-cable level but, rather, at the mechanical interface between the coil winding and the supporting structures. Hence they are generally produced at some distance from the superconducting wire, and energy reaches the superconductor only after a diffusion process through structural components and/or the insulation. For this reason, although the energy release is large, potentially reaching thousands of mJ·cm$^{-3}$, and affects large coil volumes, the time-scale for the energy release is long and thus the associated power is small. In spite of this, particular care must be applied when designing critical areas in a coil system, such as the interface between the coil winding and the mechanical structure (e.g. coil flanges and formers for solenoids), interfaces between winding parts (e.g. mating surfaces of segmented magnets), and coil interconnections (e.g. soldered joints between windings), aiming on the one hand to avoid displacements and on the other to minimize the energy release following the unavoidable deformations.

## 2.3 Electromagnetic transients

In several cases, superconductors must be either designed for pulsed operation (e.g. transformers, power cables, SMES systems) or must be able to withstand transient changes of the self and background magnetic fields (e.g. the effect of a plasma disruption in a superconducting magnet system of a thermonuclear experiment, or the tripping of normal resistive inserts in a hybrid solenoid). In any case, all superconducting magnets, whether designed for steady-state or pulsed operation, must be ramped to the operating conditions. Hence the operation of a superconductor in a magnet is always associated with more or less severe conditions with regard to the variation of the field seen by the cable.

Any field change, on the other hand, produces energy dissipation through hysteresis or coupling a.c. loss in superconducting filaments, strands or cables. The time-scale of the energy deposition in this case is governed by the dynamics of the magnetic field as well as the time constant of the induced persistent or coupling currents, and can cover a large span of characteristic time, from a few milliseconds up to quasi-steady-state heating conditions for a.c. operation. The energy deposited by a.c. loss can also vary largely, depending on the field change, its time-scale and the characteristics of the superconductor.

In well-designed cables, the a.c. loss deposited during pulsed operation or fast field transients (typical field variations of a few teslas per second, lasting 10–100 ms) ranges from a few $mJ \cdot cm^{-3}$ to a few hundreds of $mJ \cdot cm^{-3}$. Larger energy deposition is usually avoided by a small filament diameter, thus reducing the hysteresis loss, and the placement of resistive barriers in the strands and within the cable, thus reducing the coupling loss. Continuous operation with a.c. excitation results in a distributed power deposition ranging from $0.1 \ W \cdot m^{-3}$ of cable to $100 \ W \cdot m^{-3}$ for the most demanding conditions. The typical time necessary to reach regime conditions for continuous operation is 0.1–1.0 s.

## 2.4 Heat leaks

All superconducting magnets operate at temperatures in the cryogenic range. A common source of heat is therefore represented by leaks entering through the thermal insulation of the magnet, the supporting structure, current leads and instrumentation wires. Although minimized by proper design, heat inleaks cannot be completely avoided and are removed by direct or indirect cooling. Heat through the insulation and supports in normal conditions, as well as loss of cooling or degraded cooling in abnormal operating conditions, generally result in broadly distributed heat sources of potentially large deposited energy but low power and nearly steady-state characteristics. Heat leaks from instrumentation wires or current leads can lead to more localized energy inputs with, however, long time-scales of energy deposition.

The typical heat loads due to common heat leaks have a broad range of values depending on the size and operating conditions of the magnet system. The expected values for an accelerator magnet are a few watts, while for the magnetic system of a fusion reactor the heat load represented by current leads and heat leaks can amount to a few kilowatts. The corresponding power density with reference to the cable volume in the magnet ranges from 0.1 to $10 \ W \cdot m^{-3}$. Higher power densities can be reached for localized inputs, such as a high-resistance superconducting joint or a badly thermalized instrumentation wire. The characteristic time for the establishment of regime conditions for the heat load associated with current leads or heat leaks and degraded cooling is from several seconds up to several hundreds of seconds.

## 2.5 Other sources

One additional source of energy in superconducting magnet systems used either in nuclear physics experiments or thermonuclear reactors is the heat deposited by nuclear interactions during normal and abnormal conditions. One of the most severe examples is the nuclear shower following a partial or complete loss of beam in the steering magnets of a particle accelerator, leading to energy deposition of a few mJ·cm$^{-3}$ to a few tens of mJ·cm$^{-3}$ over short periods of time, in the millisecond range, and localized over a length of a few metres, which is small compared to the overall accelerator size, but which still affects large volume portions of a magnet.

Superconducting magnet systems for thermonuclear reactors sustain a neutron flux generated by the fusion reaction. As for the continuous heat loads discussed in the previous section, although the nuclear heat can be large, the heat is deposited over long pulses, several seconds in duration, and has thus nearly steady-state characteristics. The corresponding power density ranges from 10 to 100 Wm$^{-3}$.

## 2.6 A summary of the perturbation spectrum

Figure 5 shows a summary of the energy that can be deposited by the various mechanisms discussed above in a superconducting magnet as a function of the characteristic time for energy deposition, the *perturbation spectrum*. The values reported are intended to represent guidelines for the typical orders of magnitude, and do not necessarily apply to a specific magnet system; nor do they all appear simultaneously in a given magnet systems, and they should therefore be used with caution. The representation of Fig. 7 clearly shows, however, that on the time-scale of fast energy deposition (below 1 ms) the dominating mechanisms are those associated with wire and conductor motion. Flux jumps with superconducting strands based on present-day technology are masked by the more energetic mechanical events. At intermediate time-scales, in the range a few milliseconds to a few hundreds of milliseconds, the dominant energy perturbation in pulsed magnets is generated by a.c. loss. For magnets operated in steady state in nuclear environments – for example, particle accelerators – the a.c. loss is negligible and the main concerns in this time-scale are events associated with particle showers and nuclear heat. Finally, if the magnet system operates in a steady-state condition and in a *quiet* environment – for example, MRI and laboratory magnets – the dominant events are those associated with conductor motion at the fast time-scale. For longer times, above 1 s, the dominant perturbations are generated by the steady-state heat loads (heat leaks, nuclear heat, a.c. loss) and are no longer a direct stability concern, as on this time-scale the heat is usually evacuated by the cooling system.

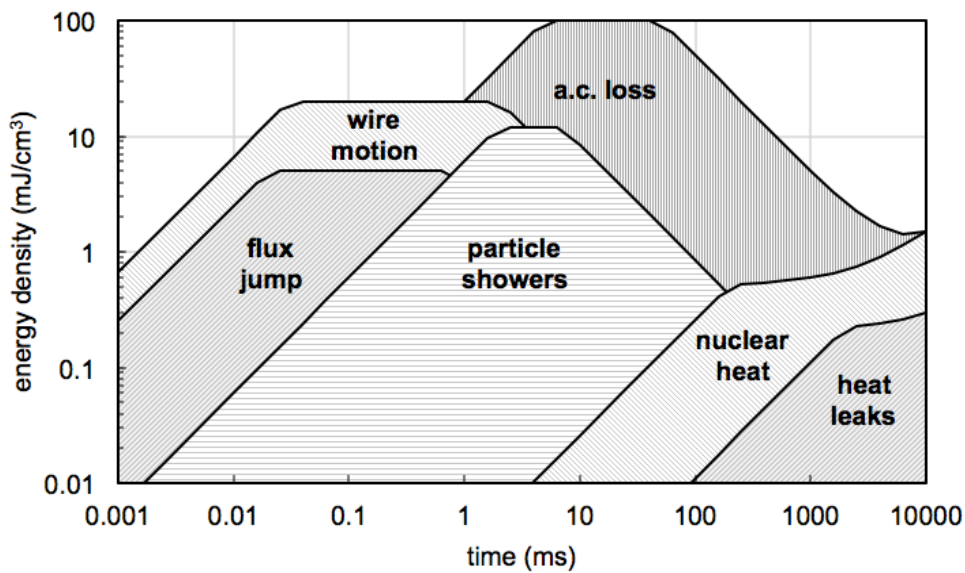

**Fig. 7:** A typical spectrum of the energy perturbations during normal operation of superconducting magnets. The energy deposited by different processes discussed here is plotted as a function of the characteristic time of energy deposition. Values are indicative and are intended for comparison of orders of magnitude.

## 3 Heat balance

The temperature of a superconducting cable changes following the energy input associated with one of the perturbations discussed in the previous section. The evolution of the temperature of the cable is governed by a transient heat balance containing, in the most general case:

- the heat-generation term representing the external perturbation;
- the Joule heating term, which appears as soon as the superconductor exceeds the current-carrying capability;
- the heat sink associated with the enthalpy of the cable;
- heat conduction along the cable and across the winding; and
- heat exchange with a coolant, with a possibly limited heat capacity, and either stagnant or flowing along the cable.

This situation is schematically depicted in Fig. 8.

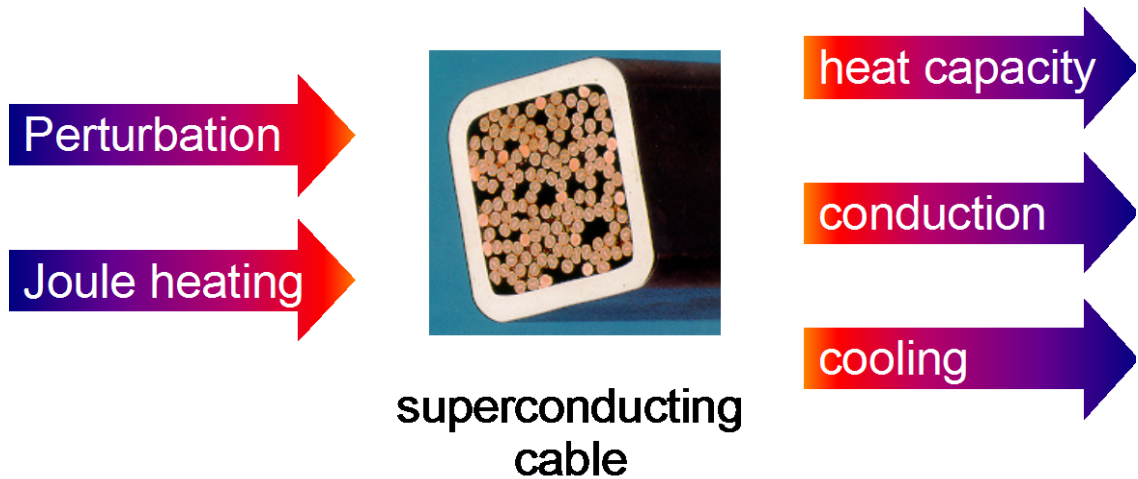

**Fig. 8:** Terms contributing to the heat balance governing the evolution of the temperature of a superconductor following the energy input associated with one of the perturbation sources discussed.

The specific form of the heat balance depends on the details of the cable considered. Accordingly, the temperature evolution, and in the final analysis the stability of the superconductor, differs for different cables. The governing equation is relatively simple in the case when the conductor can be approximated as an adiabatic strand, in which case the conduction and cooling terms disappear and analytical treatment is appropriate. On the other hand, in the case of a cable as used for large-scale applications cooled by a forced flow of helium, the terms of the equation can have a complex mathematical structure and the solution of the heat balance requires extensive numerical treatment.

In order to the discuss the main concepts necessary in order to understand the most important features of superconductor stability, we focus on the ideal case of a superconducting cable operating at a current $I_{op}$ in a background magnetic field $B_{op}$ and with an initial temperature $T_{op}$. The external energy input has a power density $q'''_{ext}$ with reference to the whole cable cross-section $A$, consisting of a superconductor portion $A_{sc}$ and a stabilizer, a low-resistance material the function of which will be discussed later, of cross-section $A_{st}$:

$$A = A_{sc} + A_{st}. \qquad (6)$$

Although other materials such as resistive barriers or structural materials could be included, we will neglect them for the moment, for the sake of simplicity. The cable temperature $T$ evolves following the perturbation, but is assumed to be uniform across the cable transverse dimension. In addition, we make the hypothesis of uniform current distribution among the strands of the cable. The cable is cooled by a helium bath at temperature $T_{he}$. Once again, for the sake of simplicity, the helium bath is considered as stagnant and with an infinite heat capacity, so that $T_{he}$ is constant. The heat exchange between the cable and the helium takes place over a *wetted perimeter w* and is characterized by a heat transfer coefficient $h$.

Under these conditions, the temperature of the cable evolves in accordance with the following equation, which is obtained from the balance of heat sources and heat sinks:

$$C\frac{\partial T}{\partial t} = q'''_{ext} + q'''_J + \frac{\partial}{\partial x}\left(k\frac{\partial T}{\partial x}\right) - \frac{wh}{A}(T - T_{he}). \tag{7}$$

In the above balance, $C$ and $k$ are the cable volumetric heat capacity and the thermal conductivity, defined as a weighted average of the properties of the superconductor and of the stabilizer (subscripts 'sc' and 'st', respectively):

$$C = \frac{A_{sc}\rho_{sc}c_{sc} + A_{st}\rho_{st}c_{st}}{A}, \tag{8}$$

$$k = \frac{A_{sc}k_{sc} + A_{st}k_{st}}{A}, \tag{9}$$

where $c_{sc}$ and $c_{st}$ are the specific heats, $\rho_{sc}$ and $\rho_{sc}$ are the mass densities, and $k_{sc}$ and $k_{st}$ are the thermal conductivities. The term $q'''_J$ stands for the heat per unit cable volume generated when the superconducting material is driven above critical conditions, in the so-called *current sharing* regime. This is the topic of the following section, and for the moment we simply assume that $q'''_J$ is zero when the superconductor is operating below critical conditions, and different from zero above.

A sample of a typical temperature evolution following an energy perturbation in a superconducting cable as described by the heat balance equation (Eq. (7)) is shown in Fig. 9. The external perturbation inducing the thermal transient is assumed to deposit its energy over a time of 10 ms. Without entering into the details of the results reported in the figure, we note that initially the temperature of the superconductor increases sharply as a consequence of the heat input provided by the perturbation. After the end of the energy pulse, the temperature initially drops under the effect of the heat conduction and cooling heat fluxes. If the conduction and cooling are sufficient, the cable recovers the superconducting state and eventually returns to stable operating conditions; that is, it follows the curve marked as *recovery* in the plot of Fig. 9. If the cooling is insufficient, after the drop the cable temperature starts to rise again under the predominant contribution of Joule heating, eventually leading to the quench shown by the curve indicated as *thermal runaway* in Fig. 9.

The balance among the non-linear Joule heating and the cooling terms is extremely delicate and can be displaced by small changes in the perturbation energy. On the basis of the qualitative features discussed above, it is possible to define an *energy margin* $\Delta Q'''$ as the minimum energy density that the external source needs to provide to the cable to cause a thermal runaway. An energy input larger than $\Delta Q'''$ causes a thermal runaway, while an energy input smaller than $\Delta Q'''$ leads to a recovery. For consistency with our discussion on the perturbation spectrum, we measure the energy margin in units of [mJ·cm$^{-3}$]. In the literature, the energy margin is often also quoted as a *stability margin*, with the same definition as above. For perturbations of known and limited distribution in space, it can be useful to refer to the *minimum quench energy*, $\Delta Q$ (MQE), which corresponds to the integral in space of the energy margin and is thus measured in units of [J].

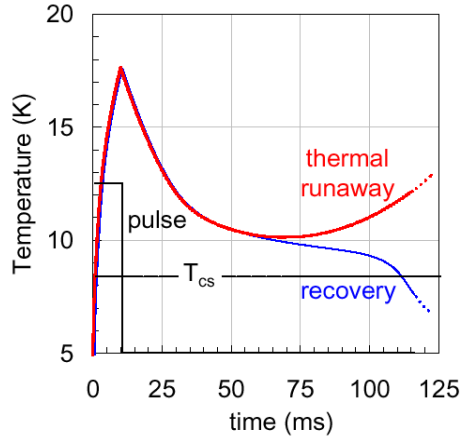

**Fig. 9:** The qualitative evolution of the temperature in a superconducting cable for an energy perturbation just below and just above the energy margin. The former leads to a recovery, while the latter results in a thermal runaway.

## 4 Current sharing

An ideal superconductor becomes resistive if one of the three parameters current density, temperature, or magnetic field exceeds the boundary of the critical surface $J_c(B,T)$, and in these conditions the current flow is associated with resistance and Joule heating in the cable. This is schematically shown in Fig. 10, which presents the temperature dependence of the critical current $I_c$ of a superconducting cable, defined as the product of the critical current density and the superconductor cross-section:

$$I_c = J_c A_{sc}. \tag{10}$$

When the cable operates at a temperature $T_{op}$ below the critical current, the material is superconducting. This situation can also be ideally maintained at a temperature above $T_{op}$ provided that the operating current is still smaller than the critical current. The temperature at which the critical current equals the operating current is called the *current sharing temperature*, $T_{cs}$. Above $T_{cs}$, the superconductor develops a resistance and the current flow is associated with Joule heating. The difference between the current sharing temperature and the operating temperature is often referred to as the *temperature margin*, $\Delta T$.

Technical low- and high-$T_c$ superconductors have a high resistivity in the normal state compared to normal conductors in cryogenic conditions. As an example, Nb–Ti has a normal state resistivity $\eta_{sc}$ of the order of $6.5 \times 10^{-7}$ $\Omega \cdot$m, while copper and aluminium have typical low-temperature resistivities $\eta_{st}$ of a few $10^{-11}$ $\Omega \cdot$m to $10^{-10}$ $\Omega \cdot$m, depending on the degree of purity and on the background magnetic field. As is intuitive, and is discussed later in detail, a decrease in heat generation is always beneficial for stability. For this reason, in addition to protection and manufacturing considerations, superconducting strands are nearly always built as a composite containing both superconducting material and a normal conducting *stabilizer* material with low resistivity in intimate mechanical, thermal and electrical contact.

The normal conducting stabilizer acts as a low-resistance shunt in the case of transition of the superconductor to the normal state, in what is called the *current sharing* process. A good representation of current sharing can be achieved considering the superconductor and the stabilizer as parallel resistors, as shown schematically in Fig. 11. For the moment, we ignore the details of the current distribution within the strand cross-section, a safe hypothesis in the time-scales of interest for stability for common strands, but insufficient when dealing with super-stabilized conductors, as will be discussed in a later section.

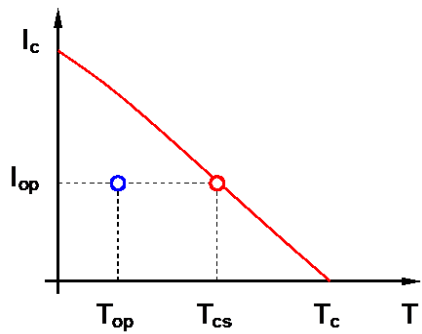

**Fig. 10:** The dependence of the critical current on temperature and the definition of the current sharing temperature.

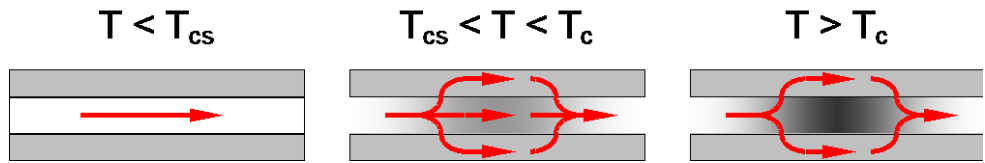

**Fig. 11:** Current sharing in a composite strand consisting of a superconductor and a normal conducting material. The situation is shown for operation below the critical surface of the superconductor (left), in current sharing conditions, at a temperature between the current sharing temperature and the critical temperature (centre), and above the critical temperature (right).

For an operating temperature below $T_{cs}$, the whole operating current $I_{op}$ flows in the superconductor, with zero resistance and no Joule heating. For an operating temperature between $T_{cs}$ and $T_c$, the superconductor develops a longitudinal resistive voltage. Under this voltage, a part of the current is transferred from the superconductor to the stabilizer. The current in the stabilizer also produces a longitudinal resistive voltage, and in equilibrium conditions this is equal to the voltage in the superconductor. The amount of current transferred depends on the electrical characteristics of the superconductor (in normal state) and of the stabilizer. As discussed above, the normal state resistivity of the superconductor is much larger than that of the stabilizer. This corresponds to the voltage–current characteristic schematically shown in Fig. 12, with zero resistivity up to the critical current and infinite resistivity above.

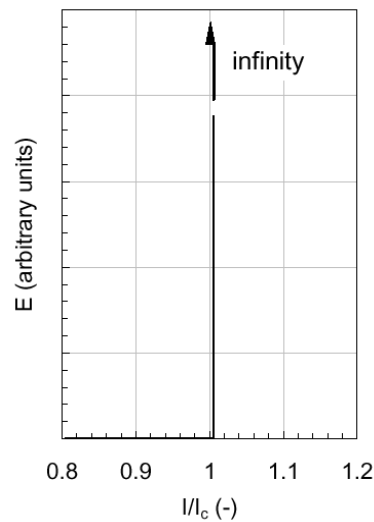

**Fig. 12:** The voltage–current characteristic of an ideal superconductor. The resistivity is large in the normal state, for a ratio of operating current to critical current above 1. In these conditions, the longitudinal electric field has an ideally infinite value.

In this condition, the current transferred to the stabilizer is exactly the current in excess of the critical current, or

$$I_{st} = I_{op} - I_c, \qquad (11)$$

while the superconductor carries the current $I_c$. The longitudinal electric field $E$ in the stabilizer (and in the superconductor) is given by

$$E = I_{st} \frac{\eta_{st}}{A_{st}}, \qquad (12)$$

and the Joule heat power density in the cable can be calculated as follows:

$$q'''_J = \frac{EI_{st} + EI_{sc}}{A} = \frac{EI_{op}}{A} = \frac{\eta_{st}}{A_{st}} \frac{I_{op}(I_{op} - I_c)}{A}. \qquad (13)$$

Finally, for an operating temperature above $T_c$, the critical current of the superconductor is zero and the whole current flows in the stabilizer. In this case, the longitudinal electric field is

$$E = I_{op} \frac{\eta_{st}}{A_{st}} \qquad (14)$$

and the Joule heat power density is given by

$$q'''_{J\max} = \frac{\eta_{st}}{A_{st}} \frac{I_{op}^2}{A}. \qquad (15)$$

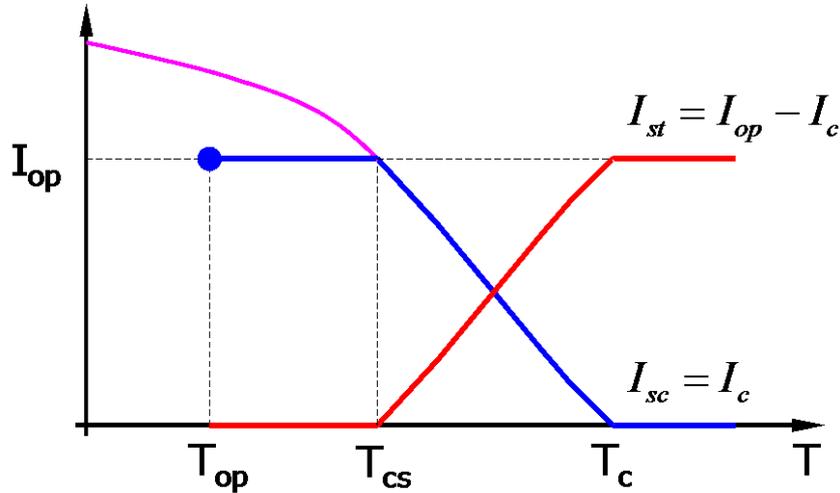

**Fig. 13:** Current sharing between the superconductor and the stabilizer as a consequence of a temperature change

The situation described above is shown schematically in Fig. 13. We see by direct comparison of Eqs. (13) and (15) that the maximum Joule heating is reached in the condition $T > T_c$, as expected. So far, we have made no assumptions with regard to the temperature dependence of the critical current, and the expressions derived above are quite general. It is, however, customary to take a line approximation for the $I_c(T)$ dependence, writing that

$$I_c \approx I_{op} \frac{T_c - T}{T_c - T_{cs}}. \qquad (16)$$

In this case, we can write explicitly the temperature dependence of the longitudinal electric field in the current sharing regime:

$$E = I_{op} \frac{T - T_{cs}}{T_c - T_{cs}} \frac{\eta_{st}}{A_{st}}, \qquad (17)$$

which is also a linear function of temperature, rising from zero at the current sharing temperature $T_{cs}$ to its maximum at the critical temperature $T_c$. The Joule heat power density is given by

$$q'''_J = \begin{cases} 0 & \text{for } T < T_{cs}, \\ q'''_{J\max}\left(\dfrac{T - T_{cs}}{T_c - T_{op}}\right) & \text{for } T_{cs} < T < T_c, \\ q'''_{J\max} & \text{for } T > T_c, \end{cases} \qquad (18)$$

where the maximum Joule heating power density is defined as in Eq. (15). The functional dependence reported in Eq. (18) is represented schematically in Fig. 14. Note the non-intuitive linear increase of the Joule heating power between $T_{cs}$ and $T_c$.

The ideal situation discussed so far provides a fair description of most situations, with a simple *caveat* on the fact that the stabilizer resistivity entering the expressions above usually has a strong dependence on the magnetic field below 20 K and on temperature above 20 K. Once this dependence is taken into account, the result of Eqs. (15) and (18) can be used for most practical cases.

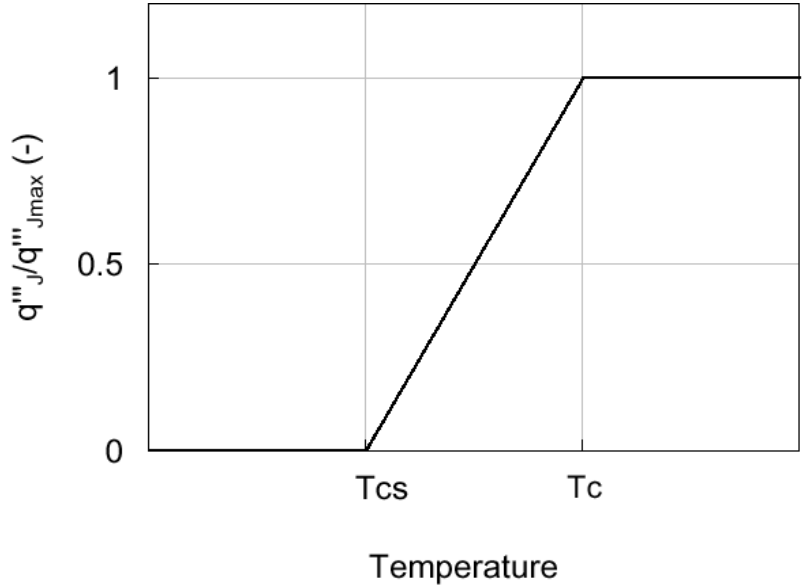

**Fig. 14:** The Joule heating power as a function of temperature, normalized to the maximum value reached for operation above $T_c$.

In particular situations – for example, when dealing with operating conditions a few per cent away from the critical current, or when considering large cables manufactured with from several hundreds to a thousand strands – it is necessary to correct the above result to achieve an accurate description. In these particular cases, it is no longer sufficient to use the asymptotic voltage–current characteristic shown in Fig. 12. A better description for the resistive transition of the superconductor is achieved using, for instance, the power-law approximation given by

$$E = E_0 \left(\frac{I_{sc}}{I_c}\right)^n, \qquad (19)$$

where $E_0$ is the electric field used as a criterion for the definition of the critical current and $n$ is the exponent defining the *sharpness* of the transition. The power law has the form shown in Fig. 15.

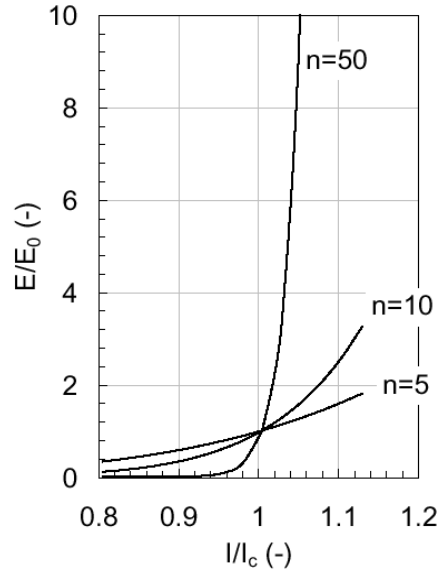

**Fig. 15:** The normalized voltage–current characteristic of a superconductor characterized by a power-law dependence of the resistive voltage on the current, with exponent $n$. The voltage–current characteristic is plotted for different values of the exponent $n$.

A low value of the exponent $n$ corresponds to a shallow and broad transition, while a high value of the exponent $n$ gives a sudden transition. Note also that a low exponent $n$ corresponds to the appearance of a resistive voltage *before* reaching the critical current. We should therefore expect that a small portion of the operating current is already transferred to the stabilizer below the critical conditions. Finally, the ideal limit used so far is achieved when the exponent $n$ tends to infinity.

The current sharing between the superconductor described by the power law and the stabilizer can be computed by equating the longitudinal electric field as done previously. The electric field in the superconductor is given by Eq. (19). For the stabilizer, Eq. (12) still holds, but we rewrite it as follows:

$$E = \left(I_{op} - I_{sc}\right)\frac{\eta_{st}}{A_{st}}. \tag{20}$$

The equilibrium condition is that

$$E_0\left(\frac{I_{sc}}{I_c}\right)^n = \left(I_{op} - I_{sc}\right)\frac{\eta_{st}}{A_{st}}, \tag{21}$$

which cannot be solved analytically for an arbitrary value of the exponent $n$. It is, however, possible to solve Eq. (21) numerically, obtaining the value of the current in the superconductor and in the stabilizer. The Joule heat power density is then computed as in Eq. (13), where this time the simplest form is obtained as follows:

$$q_J''' = \frac{EI_{st} + EI_{sc}}{A} = \frac{EI_{op}}{A}. \tag{22}$$

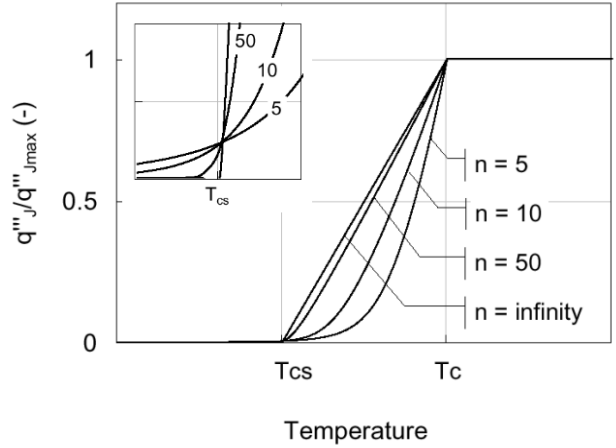

**Fig. 16:** The normalized Joule heat dependence on temperature for a superconductor with a power-law voltage–current relation. The normalized heat generation is plotted for different values of the exponent $n$ and compared to the ideal case obtained for an infinite value of $n$. The inset shows a detail of the heat generation in the vicinity of the current sharing temperature $T_{cs}$.

A sample of the numerical solution obtained for normalized heat generation is shown in Fig. 16. We note that low values of the power-law exponent $n$ correspond to less heat generation in the current sharing regime, between $T_{cs}$ and $T_c$. The implications of this fact for stability will be discussed later. On the other hand, a peculiar feature to be remarked in the case of low $n$ is that the Joule heating already starts before reaching $T_{cs}$, which is consistent with the appearance of an electric field already, below the critical conditions, as discussed earlier. In fact, early current sharing and Joule heating at low $n$ can be a limiting factor for operation at a high fraction of the critical current, and for this reason a high value of the exponent $n$ is considered an indicator of the good quality of the strand or cable. Finally, note how the linear limit given by Eq. (18) is approached when $n$ is large, as expected.

## 5    Heat transfer

With the sole exception of superfluid helium, heat transfer in cryogenic fluids has been found to be very similar to that predicted by standard thermodynamics. Proper allowance must be made for the fact that the thermophysical properties at the operating point of relevance are very different from those of room- and high-temperature coolants. Apart from this, however, the correlations available in the literature for the various room- and high-temperature heat transfer regimes are essentially also valid in cryogenic conditions with small adaptations. For low-temperature superconducting magnets two regimes are of particular relevance, namely boiling heat transfer to a stagnant bath of atmospheric pressure, saturated liquid helium (temperature around 4.2 K, pressure approximately 1 bar) and forced-flow convection of supercritical helium (temperature in the range of 4.5 K, pressure above 3 bar). For these two regimes, we will give practical correlations and typical values of heat transfer that will be used later in the discussion. In addition, in more recent years, following advancements in the technology necessary to produce superfluid helium, cooling in a bath of stagnant, sub-cooled superfluid helium at atmospheric pressure is used in several small and large-sized applications. Heat transfer in superfluid helium is peculiar and we will therefore discuss a simple approximation to this process.

### 5.1    Boiling heat transfer

Cooling of the first large-sized superconducting magnets, such as the Big European Bubble Chamber (BEBC), was achieved by submerging the magnet in a saturated bath of stagnant helium at atmospheric pressure and a temperature of 4.2 K. Heat transfer in these conditions is associated with phase transition from liquid to vapour, the boiling process. The heat transferred from the heated surface to the helium in the boiling regime has a complex but known behaviour, shown schematically in Fig. 17.

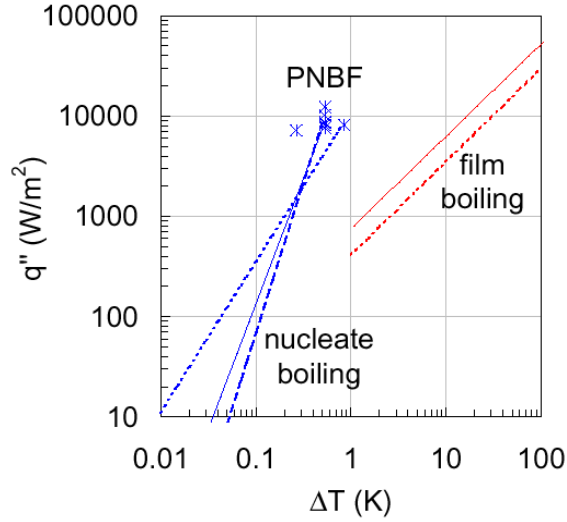

**Fig. 17:** The heat flux for boiling heat transfer regimes: the transition between nucleate and film boiling conditions is unstable.

In the figure, we report a collection of results of correlation fits to measured heat flux data as a function of the temperature difference between the heated surface and the bulk of the bath [6, 7]. If the temperature difference is small, heat transfer takes place in the nucleate boiling regime, where the heat transferred is proportional to approximately the third power of the temperature difference. For temperature differences in the range of 0.5 K, the heat transfer reaches a crisis point at which conditions become unstable. The maximum heat flux that can be reached is the Peak Nucleate Boiling Flux (PNBF), which depends on the nature of the heated surface. The material, surface roughness, surface coating, and surface orientation with respect to gravity can affect the heat transfer and PNBF by a factor of 2 [8]. The value of the PNBF for helium, however, has never been found to exceed 10 kW·m$^{-2}$. If the heat flux is increased above the PNBF, the surface becomes covers with a film of vapour and the heat transfer degrades. This regime is called film boiling and is characterized by a linear dependence of the heat flux on the temperature difference. The transition from nucleate to film boiling and back is hysteretic, takes place at constant heat flux and is characterized by a sudden jump in the temperature difference at the surface.

From the typical data of Fig. 17, it is possible to calculate an effective heat transfer coefficient, $h$, as the ratio of the heat flux to the temperature difference $\Delta T$ between the heated surface at temperature $T_s$ and the helium at temperature $T_{he}$. This has been done below, leading to the following approximate expressions:

$$h_{\text{Nucleate}} = 405 + 37 \times 10^3 \left(T_s - T_{he}\right)^{1.4}, \qquad (23)$$

$$h_{\text{Film}} = 592 \left(T_s - T_{he}\right)^{-0.077}, \qquad (24)$$

which hold for saturated helium at atmospheric pressure. The result of the above expressions is shown in the compilation of Fig. 18, together with other heat transfer mechanisms discussed later. The nucleate boiling regime is associated with extremely high values of $h$, changing rapidly and ranging from a few hundreds of W·m$^{-2}$·K$^{-1}$ to well above 10 kW·m$^{-2}$·K$^{-1}$ for a temperature difference of 0.5 K. As soon as the PNBF is reached, however, the equivalent heat transfer drops to a constant value, of the order of 400–600 W·m$^{-2}$·K$^{-1}$.

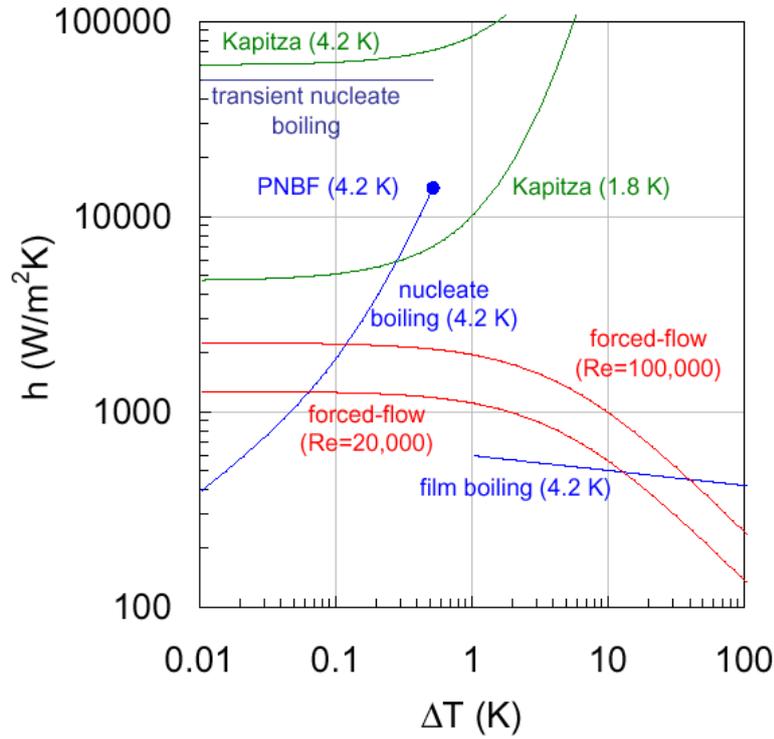

**Fig. 18:** A summary of the equivalent heat transfer coefficient for different heat transfer regimes in helium

During the transient processes of interest for stability, the heat transfer to a bath of saturated helium is substantially different with respect to the steady-state behaviour discussed above. In measurements, it is found in particular that nucleate boiling persists at much higher heat fluxes than those observed in steady-state conditions, more than one order of magnitude higher than the PNBF discussed above. This is the effect of thermal diffusion in the helium in direct contact with the heated surface [9]. The heat transfer crisis is reached in transient conditions when the helium volume affected by thermal diffusion has absorbed an energy equivalent to the latent heat of evaporation, at which point a transition to film boiling takes place. Also, the equivalent heat transfer coefficient during transient boiling can reach extremely high values, of the order of 50 kW·m$^{-2}$·K$^{-1}$, probably limited by Kapitza resistance at the solid wall [9].

### 5.2 Forced flow

Steady-state heat transfer to a turbulent forced flow of supercritical helium appears to be well approximated by a correlation of the Dittus–Boelter form, as shown by Yaskin [10] and Giarratano [11]. A best fit of the available data is obtained with the following expression, which includes a correction for large temperature gradients at the wetted surface:

$$h_{DB} = 0.0259 \frac{K_{He}}{D_h} Re^{0.8} Pr^{0.4} \left( \frac{T_{he}}{T_s} \right)^{0.716}. \qquad (25)$$

The steady-state forced-flow heat transfer for a common Reynolds number in cable cooling pipes (Reynolds in the range of a few $10^4$ to $10^5$) is usually of the order of 1000 W·m$^{-2}$·K$^{-1}$. A large temperature difference, of several degrees kelvin, between the heated surface and the bulk causes an appreciable degradation of the above values.

As for the boiling conditions, strong variations of the heat transfer are observed during fast transients. Giarratano [12] and Bloem [13] measured the transient heat transfer to a forced flow of supercritical helium in dedicated measurements on short test sections (see Fig. 19).

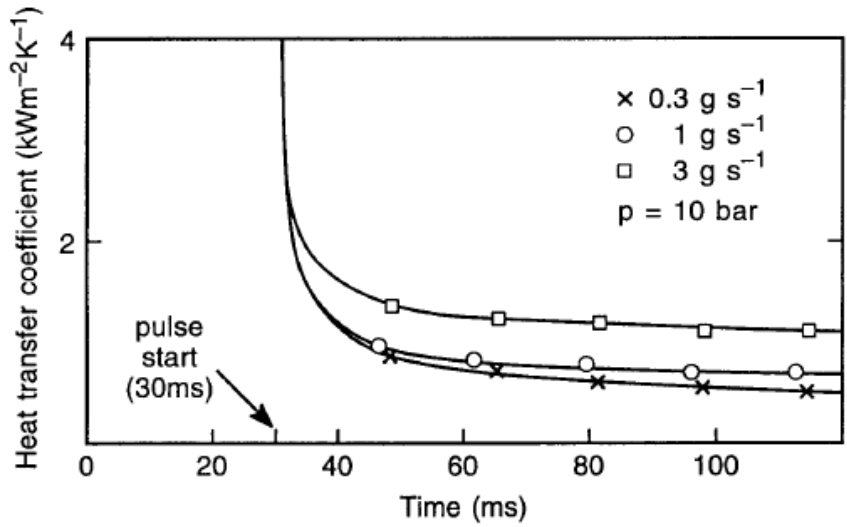

**Fig. 19:** The transient heat transfer coefficient in supercritical helium, measured by Bloem [13]. (Reprinted from Ref. [13] with kind permission from Butterworth–Heinemann Journals, Elsevier Science Ltd, The Boulevard, Langford Lane, Kidlington OX5 1GB, UK.)

The experiments showed an initial peak in the heat transfer coefficient at early times, below 1 ms. At later times, in the range of some milliseconds to about a hundred milliseconds, the initial peak decreased approximately with the inverse of the square root of time. This behaviour can be explained in terms of the diffusion of heat in the thermal boundary layer. Using the analytical solution of diffusion in a semi-infinite body (the helium) due to a heat flux step at the surface, the effective heat transfer coefficient can be computed as (Bloem [13])

$$h_{BLQ} = \frac{1}{2}\sqrt{\frac{\pi K_{he}\rho_{he}c_{he}}{t}} , \qquad (26)$$

where $K_{he}$ is the heat conductivity of helium. The above expression is shown to fit the experimental data properly for times longer than a millisecond and until the thermal boundary layer is fully developed. At early times, Eq. (26) would tend to predict an exceedingly high heat transfer coefficient, consistent with the assumptions of the analytical calculation. In reality, the early values of $h$ are found to be limited by the Kapitza resistance [14] at the contact surface of the strand, which gives a significant contribution only when the transient heat transfer coefficient is of the order or larger than 10 kW·m$^{-2}$·K$^{-1}$ (or in the case that the wetting helium is in the superfluid state as discussed later). At later times, usually around 10–100 ms, the thermal boundary layer is fully developed and the steady-state value of $h$ is approached. An empirical expression for the heat transfer during the transient, describing the transition from transient to steady-state conditions, can be obtained as follows:

$$h = \max\{h_{BLQ}, h_{DB}\} , \qquad (27)$$

giving good agreement with the experimental results, and showing how for short pulses the heat transfer coefficient only depends on the helium state and not on the flow conditions.

During the flow transients generated by the heating-induced flow, the two processes are combined; that is, the boundary layer changes in thickness during the thermal diffusion process. Experimental measurements in these conditions, and in particular on transient heat transfer over long lengths, pose some significant problems and results are so far not available. This issue is important, as increased turbulence in the flow can contribute to the stability margin. It is not clear whether the phenomenon has a local nature or depends on the heated length and the time-scales involved in the establishment of the expulsion of helium from the normal zone.

## 5.3 Superfluid helium

Helium undergoes a quantum transition, very similar to the phenomenon of superconductivity, when it is cooled below the so-called *lambda* point – that is, 2.17 K – at ambient pressure. In this state it becomes superfluid helium, characterized by very low viscosity and an exceedingly high thermal conductivity that allows removal of heat at high rates both at the surface interface with solid materials (e.g. a superconducting strand) as well as in the fluid bulk. For this reason, superfluid helium is used as a coolant in high-performance magnetic systems based on Nb–Ti or $Nb_3Sn$, where the low operating temperature is used to boost the critical current density. An additional advantage is that the thermal conductivity of a superfluid helium bath can be used to evacuate the heat loads in the magnetic system without requiring fluid convection. The high heat transfer rate of superfluid helium can be described in relevant conditions by an internal convection of two fluid species, a normal component and a superfluid component that carries no entropy. The two species move in counter-flow to maintain the total density, thus also achieving net energy transport.

In the case of normal helium considered in the previous sections, heat transfer is controlled in most situations by the vapour film (boiling heat transfer) or by the thermal resistance of the boundary layer (convection). In superfluid helium, on the contrary, the heat fluxes that can be sustained in the bulk fluid are so large that the thermal resistance at the solid/fluid interface becomes important. The dominant mechanism for the heat resistance at the interface is the mismatch in the propagation of the phonons. This interface thermal resistance is usually called the Kapitza resistance, and is in principle present at all operating temperatures. The heat flux across the Kapitza resistance can be roughly approximated by the following expression:

$$q''_{Kapitza} = \sigma \left( T_s^n - T_{he}^n \right), \tag{28}$$

where the exponent $n$ is in the range of 3–4. The constant $\sigma$ depends on the nature and state of the material, and for a value of $n = 4$ its value can range from 200 to 400 $W \cdot m^{-2} \cdot K^{-4}$. With this choice of $n$, the equivalent heat transfer coefficient at the surface is then given by

$$h_{Kapitza} = \sigma \left( T_s^2 + T_{he}^2 \right)\left( T_s + T_{he} \right). \tag{29}$$

Typical values of the equivalent heat transfer coefficient are plotted in Fig. 18 for comparison with the other heat transfer regimes. At low temperature (below 2 K), the Kapitza resistance is relatively large, corresponding to a heat transfer coefficient in the range of 5 $kW \cdot m^{-2} \cdot K^{-1}$, and is usually the main limit for heat transfer. Already at 4.2 K, though, the equivalent heat transfer coefficient is extremely high, above 50 $kW \cdot m^{-2} \cdot K^{-1}$, thus explaining why the Kapitza resistance is generally not a limiting factor in heat transfer to normal helium.

## 6 Stabilization strategies

### 6.1 Adiabatic stabilization

The lower limit for the energy margin of a superconductor can be obtained by considering that the cable responds adiabatically to the external energy input. This is the case in the absence of a cryogenic cooling fluid and whenever the volume affected by the external energy input is so large that heat conduction at the boundary of the normal zone can be neglected. The absence of cooling by a stagnant or flowing cryogenic fluid, most frequently liquid helium, is typical of small windings, either wound from a *dry* superconductor or impregnated with organic resins. Dry or impregnated windings conductively cooled using a cryocooler as a heat sink are becoming increasingly attractive due to the appeal of cryogen-free operation. Cooling happens in this type of winding on a time-scale that is much longer than the time of relevance for discriminating between recovery and thermal runaway. Hence for these magnets, the cooling term is absent in the heat balance. In this case, the energy balance simplifies to the following:

$$C\frac{\partial T}{\partial t} = q'''_{\text{ext}} + q'''_{\text{J}}. \tag{30}$$

As expressed by Eq. (18), any excursion of the superconductor above the current sharing temperature $T_{\text{cs}}$ will cause the appearance of Joule heating, resulting in an inevitable thermal runaway.[2] In this case, the current sharing temperature provides the boundary between recovery and thermal runaway. The stability margin corresponds to the energy necessary to increase the superconductor temperature from operating conditions to $T_{\text{cs}}$. Within the approximations considered so far, this can be calculated by integrating Eq. (30), taking into account that below $T_{\text{cs}}$ the Joule heating is zero:

$$\int_0^\infty q'''_{\text{ext}} \, \text{d}t = \int_{T_{\text{op}}}^{T_{\text{cs}}} C \, \text{d}T. \tag{31}$$

The integral on the left-hand side of Eq. (31) corresponds to the energy margin, while the integral on the right-hand side is the difference in the volumetric specific enthalpy between the operating temperature and the current sharing temperature. We can thus write that for an adiabatic superconductor the energy margin is as follows:

$$\Delta Q''' = H(T_{\text{cs}}) - H(T_{\text{op}}), \tag{32}$$

with the definition

$$H(T) = \int_0^T C(T') \, \text{d}T'. \tag{33}$$

For this reason, the mechanism underlying adiabatic stability, namely the heat capacity of the superconductor, is also referred to as *enthalpy stabilization*.

In order to estimate the orders of magnitude of the energy margin in adiabatic conditions, it is necessary to examine the typical values of the heat capacity and the specific volumetric enthalpy of the typical materials present in a superconducting winding. This is done in Figs. 20 and 21, respectively. The functional dependence of the volumetric heat capacity is different among pure metals (copper, aluminium), alloys (stainless steel), superconductors (Nb–Ti and Nb$_3$Sn), and organic composites (resin and typical insulators) because of the different weights of the electronic and phonon contributions to the specific heat. However, in spite of the large range of values, a general feature exhibited by all materials is the decrease of the specific heat approaching absolute zero. The consequence is that the enthalpy difference for a given temperature margin $\Delta T$ is smaller at lower temperatures, as can be inferred by taking the volumetric specific enthalpy difference for a fixed temperature interval from Fig. 21. This effect is particularly important, nearly one order of magnitude, when considering operation in superfluid helium, at 1.8–2.0 K, as compared to operation in atmospheric pressure liquid helium, at 4.2 K.

---

[2] Strictly speaking, in the case of a practical superconductor with a voltage–current characteristic described by the power-law relation, Joule heating is always positive, although small, even below current sharing. In this case, the superconductor would never be stable, even in the absence of an external energy input. The steady-state Joule heating is, however, small and is removed by the cooling system – which is inevitably present – that acts on times much longer than the time-scales of interest for stability, but sufficient for maintaining the steady-state operating temperature.

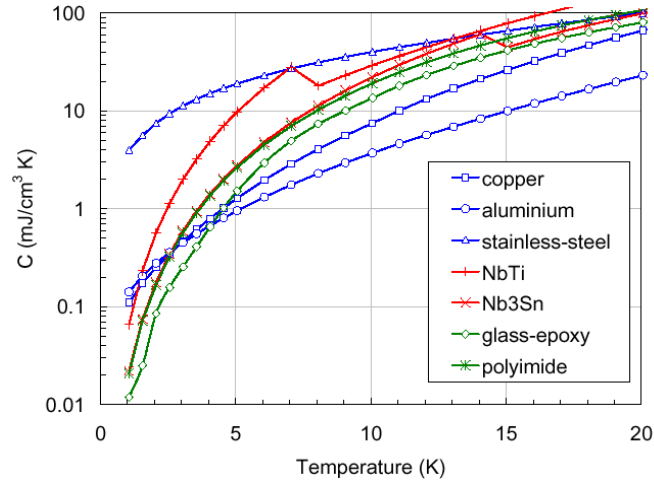

**Fig. 20:** The volumetric heat capacity for typical materials used in low-temperature superconducting magnets

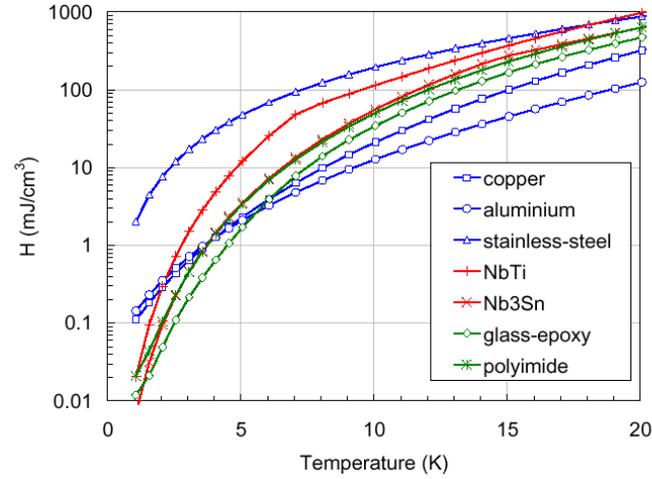

**Fig. 21:** The volumetric specific enthalpy for typical materials used in low-temperature superconducting magnets, obtained by integrating the data of Fig. 20.

The adiabatic energy margin can be estimated using the data in Fig. 21, by adding the individual contributions to the heat capacity of all materials participating in the temperature excursion. We can do this by taking the weighted average of the volumetric specific enthalpy:

$$\Delta Q''' = \sum f_i \left( H_i(T_{cs}) - H_i(T_{op}) \right), \tag{34}$$

where the $f_i$ is the materials fraction of the $i$th component characterized by the specific volumetric enthalpy $H_i$.

The energy margin in Eq. (34) depends on the operating temperature $T_{op}$ and on the operating field $B_{op}$ and current $I_{op}$ through the current sharing temperature. It is hence possible to scan the energy margin over the whole operating space for a given superconductor once the material fractions are fixed. A typical strand for low-temperature applications has a stabilizer to superconductor ratio in the range of 1.0–1.5. As we have made the assumption that the external energy input is distributed over a large volume of the winding, we also include the electrical insulation in the calculation. Typical fractions of materials for impregnated windings are in the range of 30% superconducting material ($f_{sc}$ = 0.3), 40% stabilizer ($f_{st}$ = 0.4), and 30% insulation ($f_{in}$ = 0.3).

In Fig. 22, we report the result of the calculation of Eq. (34) for Nb–Ti and Nb$_3$Sn as superconducting material operating initially at a temperature of 4.2 K. The calculation has been performed for several values of the operating fields, and is plotted as a function of the operating ratio of the critical current $I_{op}/I_c$. We can see at once the clear difference between the two materials, due to the fact that $Nb_3Sn$ has a higher critical temperature and thus exploits a region of higher heat capacity for stabilization. Also, it is clear that approaching the limits of performance of the conductor (8–10 T for Nb–Ti and 16–20 T for $Nb_3Sn$), the adiabatic energy margin for an efficient use of the superconductor (operating fraction of the critical current around 0.8) becomes extremely small (below 1 mJ·cm$^{-3}$) and the magnet will not be able to withstand even the smallest perturbation without quenching.

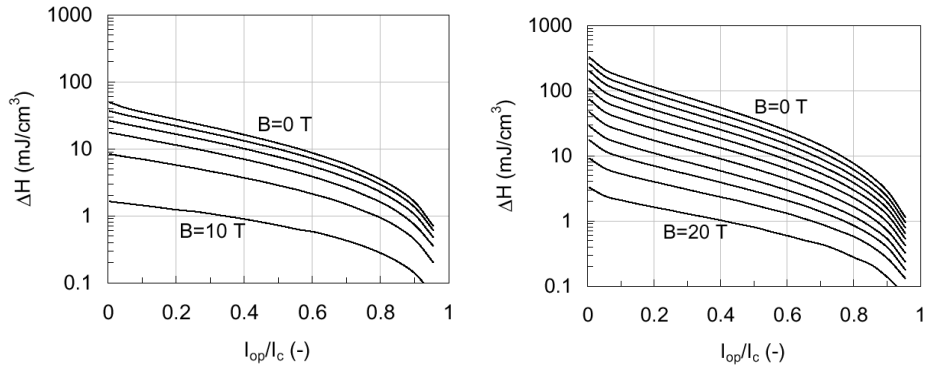

**Fig. 22:** The adiabatic energy margin for a typical winding, with 30% superconductor, 40% copper, and 30% insulation, wound either using Nb–Ti (left) or $Nb_3Sn$ (right). The calculation has been performed with an initial operating temperature of 4.2 K, for different values of the operating field (steps of 2 T between the curves) and as a function of the ratio of the operating to the critical current.

Another interesting result can be obtained by performing the above calculation for different initial operating temperatures. It is particularly instructive to examine the effect of reducing the operating temperature, which is often done to raise the critical current density in the hope of increasing the operating field of a magnet. The results of the calculation at 4.2 K, discussed in Fig. 22, are compared in Fig. 23 to the results obtained for an initial operating temperature of 1.8 K. Because a temperature change affects the critical current, to allow a direct comparison the curves are plotted as a function of the operating current density in the superconductor. We see from the comparison of the adiabatic energy margin that sub-cooling has essentially no effect at low field, as in any case the dominant contribution to the energy margin is due to the heat capacity in close proximity to $T_{cs}$.

It is only at high field (8–10 T for Nb–Ti and 16–18 T for $Nb_3Sn$) that the effect of sub-cooling becomes appreciable. However, the order of magnitude of the absolute gain in stability is at most a few mJ·cm$^{-3}$, and at best comparable with the expected mechanical and electromagnetic perturbations in a typical magnet.

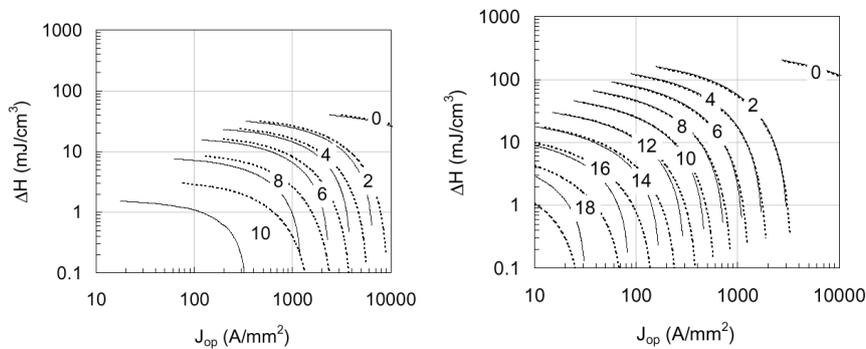

**Fig. 23:** The adiabatic energy margin for a typical winding, with 30% superconductor, 40% copper, and 30% insulation, wound either using Nb–Ti (left) or $Nb_3Sn$ (right), and computed for initial operating temperatures of 4.2 K and 1.8 K. The calculation has been performed for different values of the operating field (steps of 2 T between the curves) and as a function of the operating current density to allow direct comparison of the results.

For the above reasons, enthalpy stabilization is not sufficient to make the best use of the current-carrying potential of superconductors, and other means have been devised to cope with the perturbation spectrum, especially in large magnet systems.

### 6.1.1 *The adiabatic stability of an MRI magnet*

One of the most widespread and well-known present-day applications of superconductivity is in the magnets developed for Magnetic Resonance Imaging (MRI). These magnets are solenoids with a very good field homogeneity and a large bore, as large as 1 m in diameter and 1.5 m long, to allow full-body scans of human beings. Typical field levels in the bore of the solenoid are at present in the 1–2 T range. For cost and maintenance reasons, these magnets are built with high operating current densities, and with little or no cryogen in the winding pack. They are essentially adiabatic, and therefore they must be carefully designed to avoid training and quenches.

A typical MRI magnet winding pack is subdivided into a series of thin coaxial, possibly nested solenoids for shielding, which produce the field and correct for winding and geometrical errors. To obtain a good field homogeneity, the winding geometry must be tightly controlled and carefully maintained. In addition, the contribution from the magnetization of the superconductor must be minimized. Because of these requirements, MRI magnets are generally wound from single wires, with medium-size superconducting filaments. The winding pack is impregnated so that it forms a single rigid unit and the wires are constrained in position. Cooling is indirect, by conduction through the winding pack. The thin winding pack allows heat removal under a small temperature gradient.

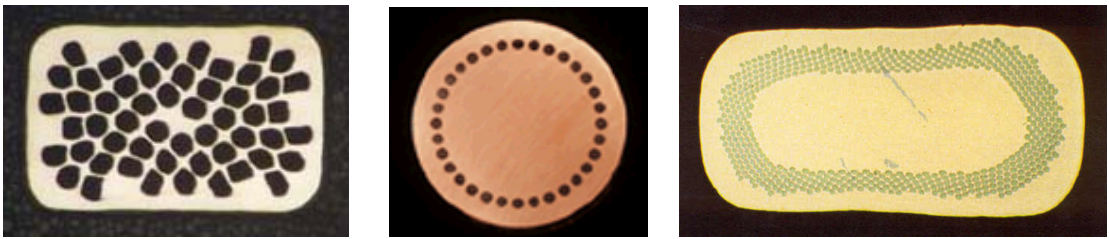

**Fig. 24:** Typical Nb–Ti strands for superconducting MRI magnets

A typical MRI magnet is wound with Nb–Ti wire, well adapted to the field range required, of the types shown in Fig. 24. The wire has a high copper/Nb–Ti ratio, of the order of 5–10, mostly for protection reasons because of the large inductance of the coil. Wires for MRI magnets are produced with a round or rectangular shape (to ease winding), and have external dimensions of the order of 1–2 mm. The Nb–Ti filaments have a typical diameter of 50 μm. At low field, they are delivered with a guaranteed critical current density in the Nb–Ti cross-section in the range of 5000 A·mm$^{-2}$. These must be compared to operating current densities in the range of 200 A·mm$^{-2}$ in the strand; that is, of the order of 1000–2000 A·mm$^{-2}$ with reference to the Nb–Ti cross-section. We can see at once that MRI magnets are designed with large operating margins to increase their reliability. Still, additional care is necessary.

In the adiabatic case, we can estimate the stability margin as the enthalpy of the wire from operating conditions to the current sharing temperature. To do this we take in account the contribution of copper, the main component in the wire, and Nb–Ti, and we use the diagram reported in Fig. 21. Assuming an operating temperature of 4.2 K, and a current sharing temperature of 7.5 K (consistent with the operating current and field range given above), the 3.3 K temperature margin corresponds to an enthalpy change of 6.14 mJ·cm$^{-3}$ for copper and 46.7 mJ·cm$^{-3}$ for Nb–Ti. The total adiabatic energy margin is obtained using the weighted sum of Eq. (34) and gives approximately 13 mJ·cm$^{-3}$. This value is larger than our estimate of the energy release due to movement of the conductor, but does not leave much contingency to cope with uncertainties in the actual temperature margin and other possible energy inputs. Resin impregnation of the winding pack, as mentioned previously, is a common practice to avoid movements and thus to minimize energy release through wire motion. Note that cracking of the impregnation resin during cool-down and energization – resins undergo a large thermal contraction from room temperature to 4.2 K, but have little tensile strength at cryogenic temperatures – can also be a source of localized energy release. This is generally avoided by filling void volumes in the winding pack with fillers and fibre cloths or ropes that increase the tensile strength.

**6.2  Cryostability**

Early superconducting coils had a wide spectrum of large perturbations, significantly above the summary presented in Fig. 5. This was either because the strands and tapes used were prone to flux jumping, or because the mechanical design was not adapted to avoid movements, slips, insulation cracks, and the associated energy releases during energization. The adiabatic energy margin, as discussed in the previous section, was by no means sufficient to accommodate the energy perturbations. A small and localized normal zone had, in addition, no chance of recovering, because the Joule heating of the superconducting material in normal state was extremely high, and therefore the coils quenched prematurely. Based on this observation, Krantowitz and Stekly [15] and Stekly and Zar [16] added a high electrical conductivity shunt backing the superconductor, a pure copper *stabilizer*, and exposed this material to a liquid helium bath of large volume and thus constant temperature. The effect was dramatic, improving the performance of the magnet and paving the way to large-sized applications of superconductivity.

This result was achieved due to two beneficial effects: on the one hand, the Joule heat generation in the case of transition was largely decreased by the stabilizer; while on the other, the superconductor was efficiently cooled through boiling heat transfer. Cryogenic stabilization, or *cryostability*, was achieved when the steady-state composite temperature that would be attained with the full operating current flowing in the stabilizer was below the critical temperature of the superconductor. In this case the initial normal zone, caused by an arbitrary energy source, would shrink and eventually disappear.

The cryostability condition can be best understood by considering the 1D heat balance of Eq. (7) again. As the cryostability condition applies independent of the length of the conductor subjected to an energy perturbation, we can neglect the heat conduction term, and we obtain the following:

$$C\frac{\partial T}{\partial t} = q'''_{\text{ext}} + q'''_{\text{J}} - \frac{wh}{A}(T - T_{\text{op}}),\tag{35}$$

where we have made use of the fact that the helium temperature remains constant, equal to the initial operating temperature. To achieve cryostability, we are seeking the condition for which, in steady state, following the end of the energy pulse, the heat generated by the normal zone is equal to or less than the heat removal at its surface. The cryostability condition is hence obtained when

$$q'''_{\text{J}} \leq \frac{wh}{A}(T - T_{\text{op}}).\tag{36}$$

For the Joule power, we take the linear approximation of Eq. (18), which reaches the maximum given by Eq. (15) when the superconductor is at the critical temperature. Assuming for the moment a constant heat transfer coefficient, corresponding to a linear increase of the heat flux with temperature, we obtain that the cryostability condition is given by

$$\frac{\eta_{\text{st}}}{wA_{\text{st}}}I_{\text{op}}^2 \leq h(T_{\text{c}} - T_{\text{op}}).\tag{37}$$

Cryostable operation is obtained when Eq. (37) is satisfied, while in the event of higher generation or lower cooling than implied by Eq. (37), the superconductor is not cryostable. To class the mode of operation, Stekly has introduced a dimensionless coefficient $\alpha$, defined as follows [16]:

$$\alpha = \frac{\eta_{st} I_{op}^2}{hwA_{st}(T_c - T_{op})}. \tag{38}$$

Operation is cryostable for $\alpha \leq 1$, and the maximum operating cryostable current, $I_{Stekly}$, is given by

$$I_{Stekly} = \sqrt{\frac{hwA_{st}(T_c - T_{op})}{\eta_{st}}}. \tag{39}$$

The cryostability condition expressed by Eqs. (37) and (38) has a simple graphical interpretation shown in Fig. 25, obtained by tracing the heat generation and the heat removal per unit of cooled conductor surface as a function of operating temperature. In the case of the curve marked '(a)' in the plot, the operation is cryostable, as at any point, the heat generation is less than the heat removal. The limiting condition is reached with the curve marked '(b)', where the heat generation exactly matches the cooling at the critical temperature. Cryostability is violated in the case of the curves marked '(c)' and '(d)', for which a sufficiently large perturbation will raise the superconductor temperature to a region where the heat generation exceeds the cooling, thus leading to an instability. Note that once the cooling condition has been selected, it is the heat generation curve that varies with the design changes, while the heat flux to the helium is a constant.

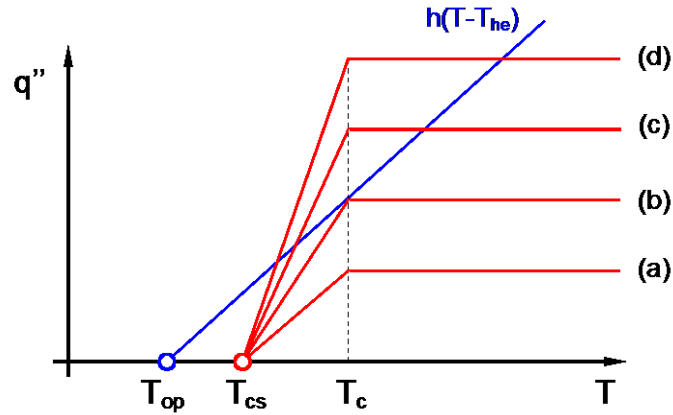

**Fig. 25:** A graphical interpretation of the cryostability conditions in the case of a constant heat transfer coefficient between the superconductor and the coolant. The graph has been obtained by plotting the heat generation and the removal per unit of a cooled conductor surface. Several heat generation values have been plotted: (a) in the cryostable region, (b) at the cryostable limit, and (c) and (d) not cryostable.

In reality, the heat transfer depends strongly on the cooling conditions, as discussed earlier. In the case of boiling helium, the typical heat flux is shown in Fig. 17. The equivalent heat transfer coefficient $h$ is non-linear, with initially high values (typically 1000–10 000 $W \cdot m^{-2} \cdot K^{-1}$) in the nucleate boiling regime, and drops to a minimum of the order of 500 $W \cdot m^{-2} \cdot K^{-1}$ at the onset of film boiling. The cryostability condition formulated above is fulfilled when, under any conditions, the heat removal exceeds the heat generation; that is, when the maximum possible heat generation is less than the minimum possible heat removal. This situation is shown in Fig. 26 for the curves marked (a) and (b). The equivalent Stekly criterion in the case of variable heat transfer is obtained taking for $h$ the minimum value along the boiling curve. As for the case of Fig. 25, curves (c) and (d) in Fig. 26 violate the cryostability condition. Note, however, that intermediate stability conditions at higher heat generation values than allowed by cryostability could exist. Taking as an example curve (c) in Fig. 26, we see that under small perturbations (a small temperature increase), the heat removal is still larger than the generation. A conductor operating in this condition would therefore recover from sufficiently small energy inputs, but it would be unstable for large enough energy depositions.

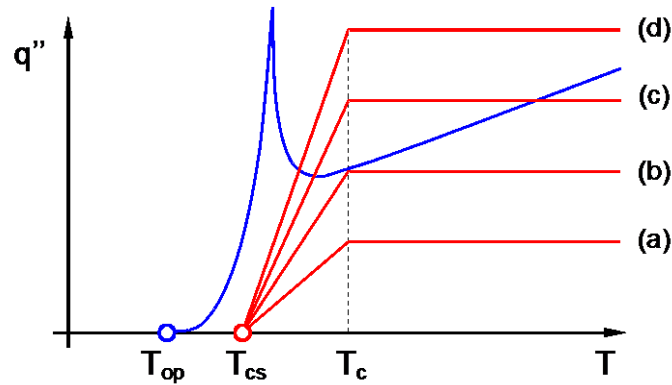

**Fig. 26:** A graphical interpretation of the cryostability conditions in the case of boiling heat transfer between the superconductor and the coolant. Cryostability is obtained for curves (a) and (b), but not for the generation curves (c) and (d).

### 6.2.1   *Cryostability: the BEBC magnet*

Cryostable magnets were among the first to be built soon after the formulation of this principle, in the early 1970s. A dramatic example was the Big European Bubble Chamber (BEBC) at CERN [17], a 4.7 m bore split solenoid with a 0.5 m gap, producing a maximum field in its centre of 3.5 T, corresponding to a maximum field at the conductor of 5.1 T, and storing an energy of 800 MJ (see Fig. 27).

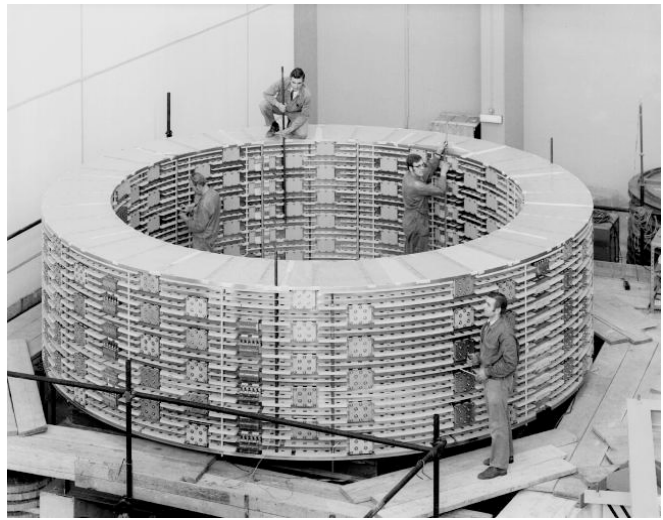

**Fig. 27:** One of the two coils forming the BEBC split-solenoid magnet at CERN

Each coil was wound in 20 pancakes out of a flat monolithic conductor with a thickness of 3 mm and a width of 61 mm, schematically shown in Fig. 28. This conductor was itself a composite, containing 200 untwisted Nb–Ti filaments with a diameter of about 200 μm in an OFHC copper matrix. The conductor had a total Nb–Ti area of approximately 6.5 mm$^2$, and about 176.5 mm$^2$ of copper cross-section. The nominal operating current of the conductor was 5700 A, corresponding to an operating current density in the composite of about 30 A·mm$^{-2}$. Adjacent conductors in a pancake were separated by insulation and by a copper spacer that allowed helium to wet the outer surface of the composite. Only one broad face of the composite was wetted (the other face was pressed against the insulation and a reinforcing steel strip), thus resulting in a wetted perimeter of 61 mm.

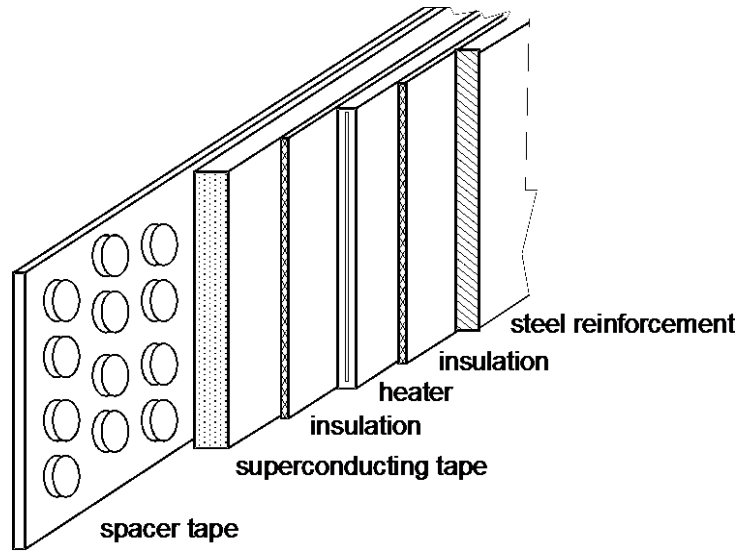

**Fig. 28:** A schematic view of the composite structure of the BEBC conductor

In order to estimate the cryostability condition, we take for boiling helium cooling, at 4.2 K, the average characteristics derived from Fig. 17 and given by Eqs. (23) and (24). The minimum heat transfer coefficient is of the order of 600 W·m$^{-2}$·K$^{-1}$. In a field of 5 T, copper has an electrical resistivity of approximately $3.4 \times 10^{-10}$ Ω·m, while Nb–Ti has a critical temperature of the order of 7.4 K. With these values, the Stekly parameter $\alpha$ is 0.55; that is, the conductor operates largely in the cryostable regime. This situation is shown in Fig. 29, which compares the heat removal and heat generation.

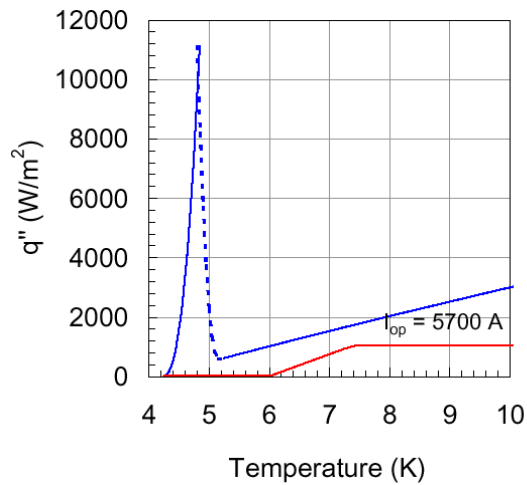

**Fig. 29:** A plot of the heat generation and heat removal per unit cooled surface of the BEBC conductor in nominal operating conditions. The conductor is cryostable, as in all possible conditions the heat removal largely exceeds the heat generation.

Indeed, the BEBC coil could be energized up to the operating current without problems, in spite of the fact that the Nb–Ti filaments were larger than the maximum size that was stable against a flux jump. In fact, because the filaments were not twisted in the composite, even larger magnetization was associated with the currents that flowed in the electromagnetically coupled filaments, excited by the field ramp. Owing to the cryostable operating regime, it was possible to suppress the large magnetization produced by these coupling currents using heaters that temporarily quenched the conductor. The conductor recovered as soon as the heaters were switched off, a rather bizarre use of cryostability.

## 6.3 Cold-end recovery

So far, all discussions have concerned a portion of superconductor long enough for all end effects to be neglected. In reality, perturbations happen often over finite lengths. We should hence expect that the conditions at the end of the resulting normal zone could help in cooling and thus provide additional stability. In these conditions, stability analysis becomes a complex matter. Nevertheless, a very simple and elegant treatment has been found by Maddock, James, and Norris [18], which identifies steady-state equilibrium conditions taking into account the effect of heat conduction along the superconductor. The situation examined is the case of a superconducting wire with a normal zone in the centre at a temperature $T_{eq}$, exchanging heat with a helium bath at constant temperature and sufficiently long that the two ends of the wire are at equilibrium temperature $T_{op}$ with the coolant. Heat is transported by conduction from the normal zone (the warm end) to the extremity (the cold end). The temperature distribution along the wire is shown in Fig. 30, together with the corresponding heat generation and removal. Only one half of the wire is plotted because of the assumed symmetry.

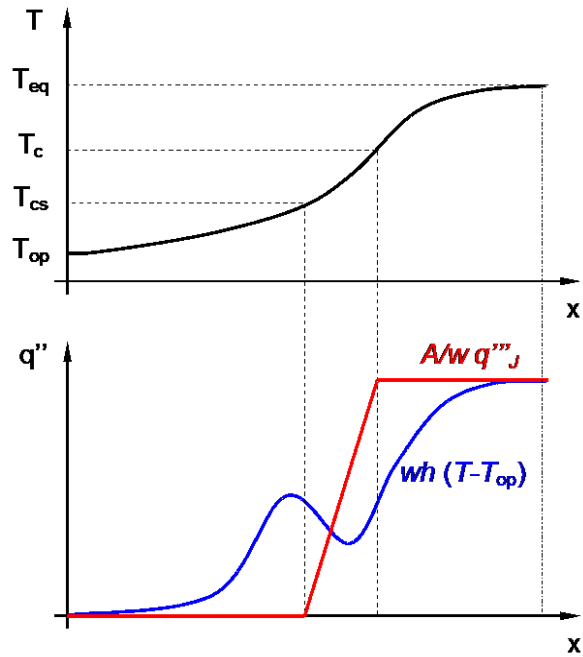

**Fig. 30:** The temperature distribution and corresponding Joule heat generation and cooling plotted as a function of the length for a superconducting wire with a normal zone in the centre. Only half of the wire length is shown because of symmetry.

The heat balance in this condition is given by Eq. (7), where in principle all terms must be retained. If we look for the equilibrium condition, however, the heat balance simplifies to

$$q_J''' + \frac{\partial}{\partial x}\left(k\frac{\partial T}{\partial x}\right) - \frac{wh}{A}(T - T_{he}) = 0 . \qquad (40)$$

Maddock and co-workers introduced a new variable $S$, defined as

$$S = k\frac{\partial T}{\partial x}, \qquad (41)$$

which represents the heat flux along the superconductor. $S$ has the property that

$$\frac{\partial S}{\partial x} = \frac{\partial S}{\partial T}\frac{\partial T}{\partial x} = \frac{S}{k}\frac{\partial S}{\partial T} . \qquad (42)$$

We can substitute the relation (41) in the steady-state heat balance of Eq. (40) and obtain the following relation:

$$k\left[\frac{wh}{A}(T-T_{he})-q_J'''\right] = S\frac{\partial S}{\partial T}. \tag{43}$$

Equation (43) can be integrated directly, yielding the following integral relation between heat generation by Joule heating and cooling:

$$\int_{T_{op}}^{T_{eq}} k\left[\frac{wh}{A}(T-T_{he})-q_J'''\right] = \int_{S_{op}}^{S_{eq}} S\, dS. \tag{44}$$

If we now make the assumption that the normal zone is sufficiently long to have reached the equilibrium condition, so that $T_{eq}$ is given by

$$T_{eq} = T_{op} + \frac{Aq_J'''}{wh}, \tag{45}$$

then the heat conduction will be zero both at the warm and at the cold end, and the relation between heat generation and cooling (for constant heat conductivity) will be simply

$$\int_{T_{op}}^{T_{eq}}\left[h(T-T_{he}) - \frac{A}{w}q_J'''\right] = 0. \tag{46}$$

Equation (46) is the so-called *equal-area theorem*, which states that for equilibrium, no net area should be enclosed between the heat generation and cooling curves plotted as a function of temperature. This very interesting result can be examined graphically on the same representation used in Figs. 25 and 26 to determine cryostability, and shown in Fig. 31 for the two cases of linear and boiling heat transfer. The point at temperature $T_{eq}$ corresponds to the intersection of the generation and cooling curves. The generation curve reported, although not cryostable, is still an equilibrium condition, as the area enclosed between generation and cooling is zero. The excess heat generation in the normal region is compensated by excess cooling in the superconducting region. Heat conduction functions as the vector of this heat from one region to the other.

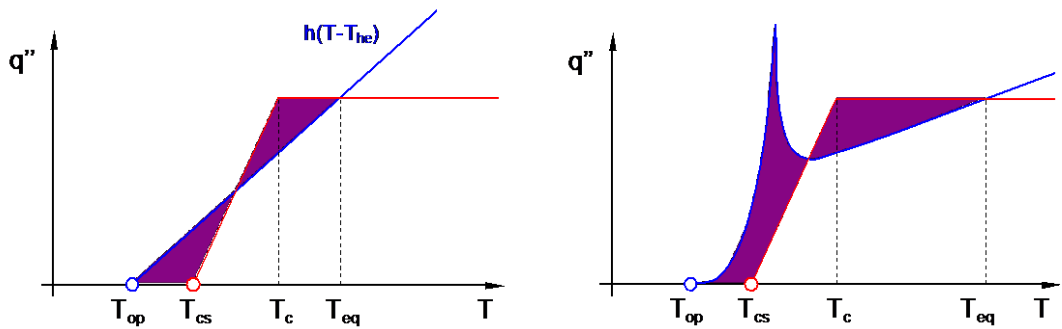

**Fig. 31:** A graphical interpretation of the equal-area condition, in the case of a constant heat transfer coefficient between the superconductor and the coolant (left) and in the case of boiling heat transfer (right). A superconductor characterized by the heat generation curves plotted is stable, although the cryostability condition is violated, as the net area enclosed between the generation and the cooling is zero.

For the linear heat transfer case, with constant $h$, it is possible to determine the value of the operating current that corresponds to the equal-area condition. To do this, we note that the equilibrium condition corresponds to the situation in which we have

$$T_{eq} - T_c = T_{cs} - T_{op}, \tag{47}$$

which derives from the similarity of the two shaded triangles in Fig. 31. From Eq. (45), we also have that the equilibrium temperature is given by

$$T_{eq} = T_{op} + \frac{\eta_{st} I_{op}^2}{wh A_{st}}. \tag{48}$$

We can combine Eqs. (47) and (48) to obtain the operating current $I_{Maddock}$ corresponding to stable operation under the equal-area condition:

$$I_{Maddock} = \sqrt{\frac{hw A_{st}\left[(T_c - T_{op}) + (T_{cs} - T_{op})\right]}{\eta_{st}}}. \tag{49}$$

As expected, this current is higher than the cryostable current $I_{Stekly}$ given in Eq. (39), implying that the superconductor can be used more effectively than when limited by cryostability. The value of the Stekly parameter for operation in the equal-area condition is given by

$$\alpha = 1 + \frac{(T_{cs} - T_{op})}{(T_c - T_{op})}, \tag{50}$$

which approaches a value of 2 for operation at a small fraction of the critical current (when $T_{cs} \approx T_c$). In reality, the temperature variation of thermal conductivity and the non-linear character of the heat transfer can cause significant variations from the above limits, with a net effect that in general is towards a higher stable operating current and a corresponding Stekly parameter.

### 6.3.1 *The equal-area condition for the BEBC magnet*

In the case already examined above for the BEBC solenoid, we have computed the value of the nominal heat generation and cooling, and verified that the BEBC magnet was operating in the cryostable regime. It is possible, using the same assumptions, to estimate the maximum operating current that could have been achieved in steady-state stable conditions as dictated by the equal-area theorem. Of course, a change in operating current would result in a simultaneous change in the magnetic field produced, of the critical properties of the Nb–Ti superconductor and of the resistivity of the copper. Linearizing the properties in the vicinity of the nominal working point, we make the following approximations:

magnetic field at the conductor: $\quad B_{op} = 0.9 \times 10^{-3} I_{op} \quad [\text{T}],$

critical temperature: $\quad T_c = 7.4\left[1 - \frac{B_{op} - 10.8}{5.1 - 10.8}\right] \quad [\text{K}],$

critical current: $\quad I_c = 13\,000\left[1 - \frac{B_{op} - 10.8}{5.1 - 10.8}\right] \quad [\text{A}],$

from which it is possible to compute the current sharing temperature in any working conditions.

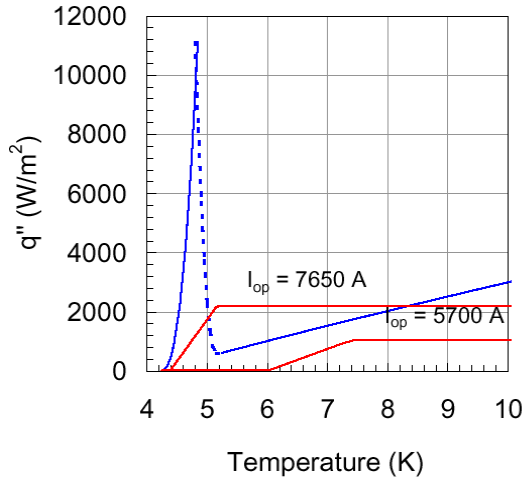

**Fig. 32:** The equal-area condition for the BEBC magnet as compared to the cryostable operating condition discussed earlier.

We can now take the same heat transfer curve as used in Fig. 29 and vary the current $I_{op}$ parametrically until the equal-area condition is reached. This situation is shown in Fig. 32. The corresponding value of the operating current is approximately 7650 A, for an operating field of 6.84 T. Ideally, this should have been the upper limit for stable steady-state operation of the BEBC magnet.

The equal-area condition guarantees stable operation at a current above the cryostability limit even in the case of a large portion of the winding going normal. This allows the designer to increase the allowable operating current density with a beneficial reduction in the amount and cost of superconductor and stabilizer used for construction of the coil. In fact, once established, the temperature profile, which obeys the equal-area theorem, is stable. If we take the example of the central point, and we refer to the generation and cooling fluxes plot of Fig. 31, a small temperature increase will cause the cooling to become larger than the generation and the conductor will evolve back to the equilibrium temperature. For a small temperature decrease, the cooling will be less than the generation and the temperature will increase back up to the balance point. In the final analysis, the equal-area theorem guarantees that the coil will not suffer from thermal runaway, whatever the energy input,[3] but, as for the cryostability condition, it does not quantify the energy margin of the conductor. Because, in addition, both the cryostability and the equal-area conditions apply to a long length of superconductor initially brought into normal conditions, the energy margin for a conductor operating in these conditions is in practice much larger than any perturbation expected during operation; that is, *infinite* from an engineering point of view.

## 6.4 Well-cooled operation of CICCs

We have assumed so far that the cooling takes place in a helium bath, providing an ideally infinite heat sink. For some applications, it is advantageous to cool the superconductor using a forced flow of helium, in which case the amount of helium available for stabilization is no longer infinite. Various superconductor configurations have been developed around this concept, of which the most successful from the stability point of view is the Cable-in-Conduit Conductor (CICC). The development of CICCs was largely motivated by the observation that cryostable pool boiling magnets (i.e. satisfying the Stekly criterion) are known to have a low operating current density, and thus a large size and cost. It was also clear, however, that large-sized magnets operating in noisy mechanical or electromagnetic environments (e.g. operating in rapidly changing magnetic fields or subject to significant stress cycles) require a minimum energy margin to withstand typical perturbations that cannot be absorbed adiabatically in the small heat capacity of the conductor.

---

[3] In reality, for large enough energy inputs, the temperature of the superconductor can become sufficiently large that the stabilizer resistivity, assumed constant so far, starts to increase sensibly. This condition, however, requires large energies and is not relevant to our discussion.

Helium is the only substance known to have a large heat capacity at low temperature. This is shown in Fig. 33, which presents the volumetric heat capacity (the product of density and specific heat at constant pressure) for different values of pressure in the supercritical regime. Comparing the values of Fig. 33 to those for solid materials in Fig. 20, we see that helium can provide a heat capacity two to three orders of magnitude larger than the solid materials in the range of 4–10 K that is typical for low-temperature superconductors. The volumetric enthalpy is shown in Fig. 34 and also demonstrates, by comparison with Fig. 21, the large heat sink that could be provided by helium.

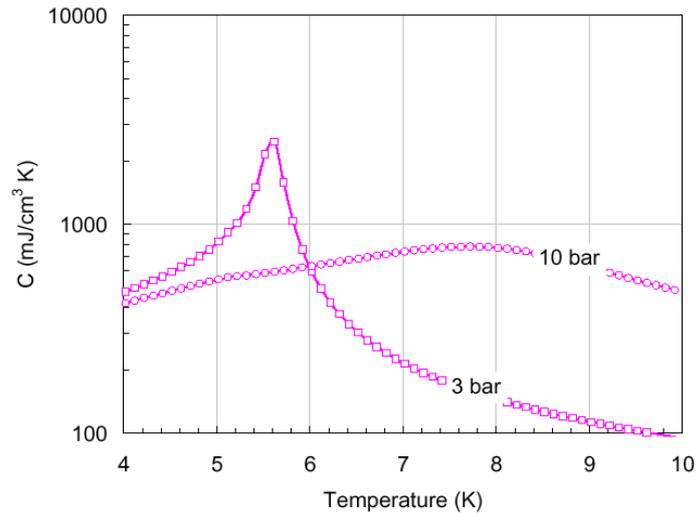

**Fig. 33:** The volumetric heat capacity of helium at different pressures: the peak corresponds to the crossing of the pseudo-critical line in the supercritical regime.

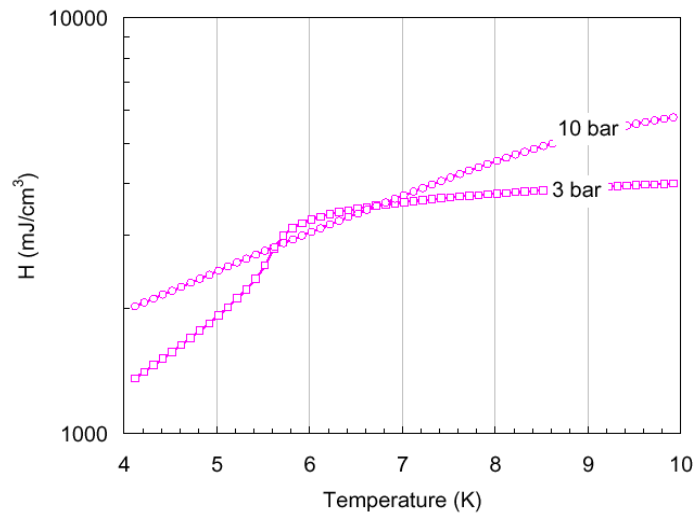

**Fig. 34:** The volumetric specific enthalpy for helium at different pressures in the supercritical regime

The key idea behind the CICC was hence to give access to the large heat sink represented by the amount of helium in the cooling circuit of the cable, thus increasing substantially the adiabatic energy margin discussed previously. At the same time, the aim was to increase the heat transfer from the superconductor to the helium, so that the cryostability limit would be pushed to higher current densities.

The CICC concept evolved from the Internally Cooled Superconductor (ICS), which had found applications in magnets of considerable size in the late 1960s and early 1970s (see, in particular, the work of Morpurgo [19]). In an ICS, the helium is all contained in the cooling pipe, very much like standard water-cooled copper conductors. The conductor can be wound and insulated using standard technology, and the magnet is stiff both mechanically and electrically, a considerable advantage for medium and large-size systems requiring, with the increasing amount of stored energy, high discharge voltages. Control of the heat transfer and cooling conditions is achieved using supercritical helium, thus avoiding the uncertainties related to a flowing two-phase fluid. A major drawback of this concept, however, was the fact that in order to achieve good heat transfer (and thus stability and a high operating current density), the helium would theoretically have to flow in the early ICS layouts at astronomical flow rates. The advantage of the increase of the wetted perimeter obtained by subdivision of the strands was already clear at the beginning of the development of the ICS (Chester [1]). Hoenig [20–22] and Dresner [23–25] developed models for the local recovery of an ICS after a sudden perturbation, where they found that for a given stability margin, the mass flow required would be proportional to the 1.5th power of the hydraulic diameter. This consideration finally brought Hoenig, Iwasa, and Montgomery [20, 21] to present the idea for the first CICC prototype, shown in Fig. 35.

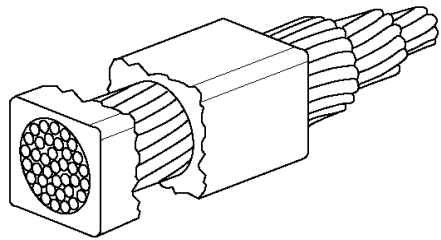

**Fig. 35:** The original concept of the CICC, as presented by Hoenig *et al.* [21]. (Reproduced from [21] by permission of Servizio Documentazione CRE–ENEA Frascati. Copyright 1975 CRE–ENEA Frascati.)

Although many variants have been considered, the basic CICC geometry has changed little since. A bundle conductor is obtained by cabling superconducting strands, with a typical diameter in the millimetre range, in several stages. The bundle is then jacketed; that is, inserted into a helium-tight conduit that provides structural support. Supercritical helium flows in the conduit, within the interstitial spaces of the cable. With the cable void fractions of about 30–40% that are commonly achieved, the channels have an effective hydraulic diameter of the order of the strand diameter, while the wetted surface is proportional to the product of the strand diameter and their number. The small hydraulic diameter ensures a high turbulence, while the large wetted surface achieves high heat transfer, so that their combination gives the known excellent heat transfer properties.

Strictly speaking, although it can satisfy the Stekly criterion (see later) a CICC cannot be considered as cryostable, because the amount of helium available for its stabilization (which represents the dominant heat capacity) is in any case limited to the volume in the local cross-section. The consequence is that a large enough energy input will always cause a quench, a behaviour that Dresner [25, 26] defines as meta-stable. Rather, the question concerns the magnitude of the energy margin $\Delta Q'''$ for a given configuration and operating condition. In the initial studies, the energy input was thought to happen suddenly, and initial experiments and theory concentrated on this assumption. Throughout this section, we will extend the definition to an arbitrary energy deposition time-scale. Finally, in spite of the fact that the cryostability concept does not apply to CICCs, we will see that the Stekly criterion, in its original form of a power balance at the strand surface, still plays a fundamental role in its stability.

Stability in CICCs is different from the theories discussed so far, for the following reasons:

- the largest heat sink providing the energy margin is the helium, and not the enthalpy of the strands themselves or conduction at the end of the heated length;

- this heat sink is limited in amount;
- finally, the helium behaves as a compressible fluid under energy inputs from the strands, implying additional feedback on the heat transfer coefficient through heating-induced flow.

As a consequence, two of the main issues in CICC stability are the heat transfer from the strand surface to the helium flow and the thermodynamic process in the limited amount of helium.

Measurements of the stability margins of CICCs started early in their history [27–31]. The original idea of reducing the necessary flow in order to obtain the desired stability margin was frustrated as soon as the first experimental data were obtained: the stability margin was largely independent of the operating mass flow, as was recognized by Hoenig [28, 29] (see the results reported in Fig. 36), and soon duplicated by Lue and Miller [31]. These results indicated that the heat transfer at the wetted surface of the strands during a temperature excursion was only weakly correlated to the steady-state mass flow and the associated boundary layer. In later experiments, Lue, Miller, and Dresner [32, 33] could observe multiple stability regions, both as a function of the operating current and of the operating mass flow (a typical stability margin showing the dual behaviour curve is shown in Fig. 37).

As discussed by Dresner [34] and Hoenig [30], during a strong thermal transient the heat transfer coefficient $h$ at the strand surface changes mainly for two reasons (see also the earlier discussion on heat transfer): (a) thermal diffusion in the boundary layer (a new thermal boundary layer is developed and thus $h$ increases compared to the steady-state value); and (b) induced flow [35] in the heated compressible helium (associated with increased turbulence and thus again an increase in $h$). The concurrence of these two effects explains the weak dependence of $\Delta Q'''$ on the steady mass flow and (at least qualitatively) the multivalued stability behaviour for different pulse powers.

The typical behaviour of the energy margin in CICCs was found through measurements to be a function of the operating current (see the vast amount of data presented in Refs. [36–41]). Such behaviour is shown schematically in Fig. 38. For a sufficiently low operating current, a region with a high stability margin – termed here, after Schultz and Minervini [42], the *well-cooled* operational regime – is observed. In this regime, the stability margin is comparable to the total heat capacity available in the local cross-section of the CICC, including both the strand material and the helium, between the operating temperature $T_{op}$ and the current sharing temperature $T_{cs}$. With increasing current, a fall in the stability margin to low values, the *ill-cooled* regime, is found. In this regime, the stability margin is lower than in the well-cooled regime by one to two orders of magnitude and depends on the type and duration of the energy perturbation.

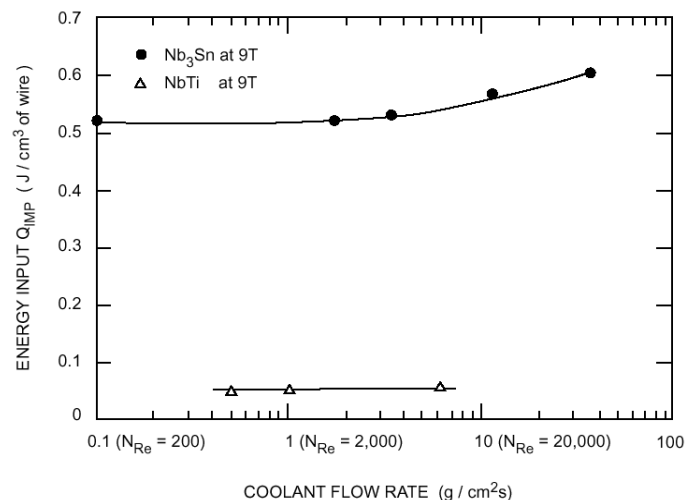

**Fig. 36:** The energy margin of a Nb–Ti CICC and a Nb$_3$Sn CICC as a function of the steady-state helium flow, measured by Hoenig *et al.* [28] (reproduced from [28] by permission of the IEEE. Copyright 1979 IEEE).

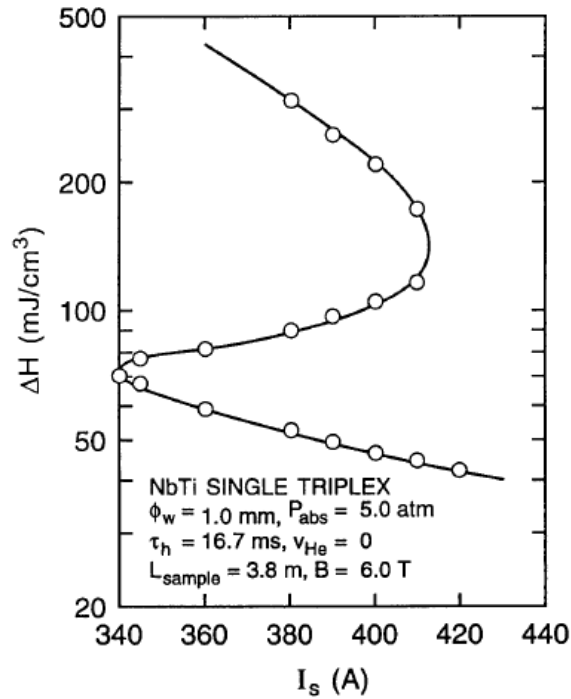

**Fig. 37:** The energy margin of a Nb–Ti CICC as a function of the operating current, measured by Lue *et al.* [33]. The experiment was performed on a single triplex CICC of length $L_{sample}$ = 3.8 m, with a strand diameter $\phi_w$ = 1 mm, under zero imposed flow ($v_{He}$) at a helium pressure of $p_{abs}$ = 5 bar. The background field was $B$ = 6 T, and resistive heating took place in $\tau_h$ = 16.7 ms. (Reproduced from [33] by permission of the IEEE. Copyright 1981 IEEE.)

The transition between the two regimes was identified by Dresner [34] to be at a *limiting* operating current, $I_{lim}$:

$$I_{lim} = \sqrt{\frac{A_{st} w h (T_c - T_{op})}{\eta_{st}}} . \tag{51}$$

The above definition of the limiting current $I_{lim}$ is equivalent to the Stekly criterion of Eq. (39). Equation (51) sets a condition necessary for recovery: the heat transfer from the strand to the helium must be larger than the Joule heat generation. This condition is satisfied for operating currents below $I_{lim}$; that is, in the well-cooled regime. On the other hand, above $I_{lim}$, in the ill-cooled regime, a normal zone will always generate more heat than it can exchange to the helium, and therefore no recovery will be possible once the strand temperature is above $T_{cs}$.

This explains the behaviour of the energy margin below and above $I_{lim}$. In the well-cooled regime, recovery is possible as long as the helium temperature is below the current sharing temperature $T_{cs}$. Therefore the energy margin is of the order of the total heat sinks in the cable cross-section between the operating temperature $T_{op}$ and $T_{cs}$, obviously including the helium. In the ill-cooled regime, an unstable situation is reached as soon as the strands are current sharing, and therefore the energy margin is of the order of the heat capacity of the strands between $T_{op}$ and $T_{cs}$ plus the energy that can be transferred to the helium during the pulse. In practical cases, the heat capacity of the helium in the cross-section of a CICC is the dominant heat sink by two orders of magnitude and more, and this explains the fall in the stability margin above $I_{lim}$.

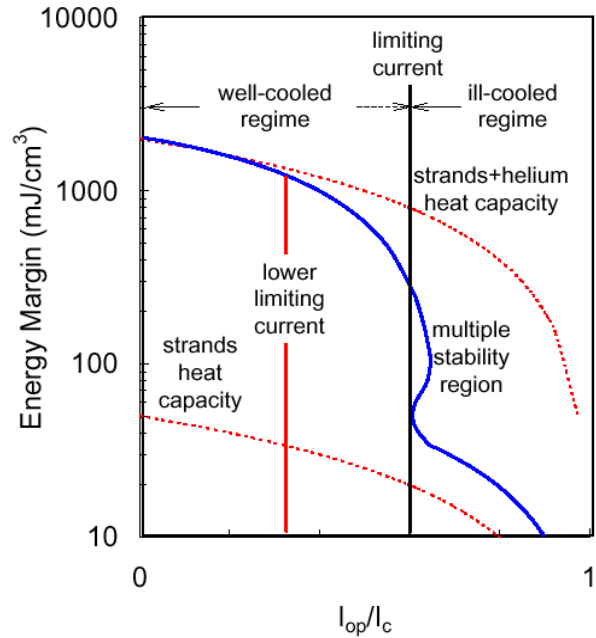

**Fig. 38:** The schematic behaviour of the stability margin as a function of the cable operating current

The transition between the well-cooled and ill-cooled regimes happens in reality as a gradual fall from the maximum heat sink values to the lower limit (Miller [39]). Defining the limiting fraction $i_{lim}$ of the critical current $I_c$ as $i_{lim} = I_{lim}/I_c$, the typical extension of this fall is of the order of $(i_{lim})^{1/2}$. An intuitive explanation for this fall can be given, again using the power balance at the strand surface. For the derivation of Eq. (51), it was assumed that the helium has a constant temperature $T_{op}$. In reality, during the transient, the helium temperature must increase as energy is absorbed, so that the power balance is displaced; that is, power can be transferred only under a reduced temperature difference between the strand and the helium. Two limiting cases can be defined. The first is the ideal condition of the helium at constant temperature, giving the limiting current of Eq. (51) – for which, however, the energy absorption in the helium is negligible. Operation at (and above) $I_{lim}$ is necessarily associated with a stability margin at the lower limit – the ill-cooled value. The second limiting case is found when the Joule heat production can be removed even when the helium temperature has increased up to $T_{cs}$. This second case is obtained for a current of (and below)

$$I_{lim}^{low} = \sqrt{\frac{A_{st} wh (T_c - T_{cs})}{\eta_{st}}}, \quad (52)$$

which we call the *lower limiting current*, by analogy to Eq. (51) and due to the fact that $I_{lim}^{low}$ is always less than $I_{lim}$. For operation at (and below) $I_{lim}^{low}$, the full heat sink can be used for stabilization and the stability margin is at the upper limit – the well-cooled value. Between the two values $I_{lim}$ and $I_{lim}^{low}$, the stability margin falls gradually, sometimes showing a multiple stability region in the close vicinity of $I_{lim}$. The multiple stability region extends over a small area that is not of interest for the safe design of a stable CICC. Therefore this feature is usually neglected.

The dependence of the stability margin on the background field $B$ is rather obviously explained by the influence on the critical and current sharing temperatures. A higher $B$ causes a drop both in the limiting current (through a decrease of $T_c$ and an increase of $\eta_{st}$) and in the energy margin (through a decrease in $T_{cs}$). Therefore, as expected, $\Delta E$ drops as the field increases. An interesting feature, however, is that the limiting current only decreases with $T_c^{1/2}$; that is, with a dependence on $B$ weaker than that of the critical current. At large enough $B$, we will always have that $I_{lim}$ is larger than $I_c$ and the cable will reach the critical current in well-cooled conditions.

The stability margin depends on the duration of the heating pulse, as shown experimentally by Miller *et al.* [31] and reported in Fig. 39. A change in the heating duration for a given energy input corresponds to a change in the energy deposition power. In the well-cooled regime – that is, for low operating currents in Fig. 39 – the heat balance at the end of the pulse is in any case favourable to recovery, and therefore the energy margin does not show any significant dependence on the pulse length. When the conductor is in the ill-cooled regime, the power removal capability is limited. For short heating pulse durations, the heating power increases and conductor reaches $T_{cs}$ faster than for lower powers, corresponding to longer heating durations. Therefore the energy margin increases at increasing pulse length until it becomes comparable to the total heat capacity (as in the well-cooled regime). This effect is partially balanced for very fast pulses, because the heat transfer coefficient can exhibit very high values (see earlier discussion) that could shift the well-cooled/ill-cooled transition at higher transport currents, and thus in principle higher energy margins should be expected in this range. However, the high input powers in this duration range tend to heat the conductor above 20 K, into a temperature range where the stabilizer resistivity grows quickly and the power balance is thus strongly influenced. This effect causes saturation of the energy margin for extremely fast pulses (well below 1 ms in duration).

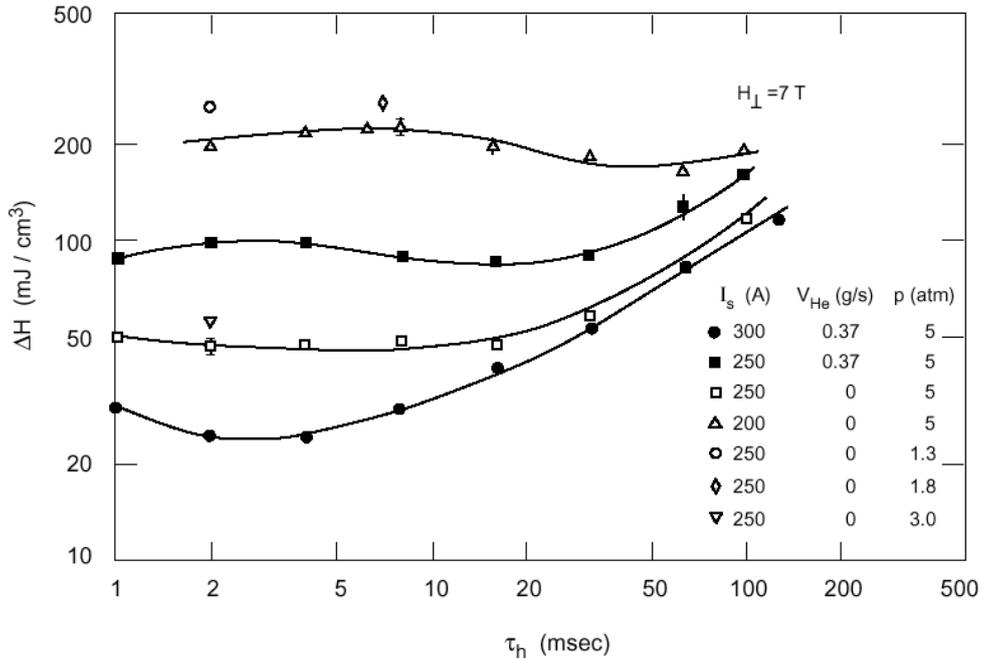

**Fig. 39:** The dependence of the stability margin for a CICC (indicated on this plot as $\Delta H$) on the heating time-scale ($\tau_h$), as measured by Miller *et al.* [31]. The parameters varied in the experiment, indicated in the inset, are the transport current in the sample, $I_s$, the helium flow velocity, $v_{He}$, and the helium pressure, $p$. (Reproduced from [31] by permission of the IEEE. Copyright 1979 IEEE.)

The dependence on the remaining operating conditions, typically the operating temperature and pressure, is not easily quantified. The reason is that the helium heat capacity in the vicinity of the usual operational regimes (operating pressure $p_{op}$ of the order of 3–10 bar and operating temperature $T_{op}$ ~ 4–6 K) varies strongly with both $p_{op}$ and $T_{op}$. This affects both the heat sink and the heat transfer coefficient (through its transient components). An increasing temperature margin under constant operating pressure gives a higher $\Delta E$. But a simultaneous variation of $p_{op}$ and $T_{op}$, under a constant temperature margin, can produce a large variation (typically of the order of a factor of 2 in the range given above) in $\Delta E$ (Miller [39]).

A mention must be made of the case in which the operating point is in the superfluid helium (He-II) range. The main difference compared to operation in He-I is the high heat transfer capability associated with superfluid helium. The presence of He-II has thus two effects. First, the power balance at the strand surface is drastically changed, being displaced towards the well-cooled condition. In addition, a significant heat flux leaks at the end of the heated region, thus making available a larger heat sink than the volume strictly contained in the heated region only. As an example, Lottin and Miller [41] measured the stability margin of a 2 m long conductor in an operating temperature range from 1.8 to 4.2 K. For this length the end effects are small, so that the experiment is a good basis to show the influence of the surface heat transfer.

The stability margin in the case of He-II operation behaves at low current in a way similar to what would be expected in the case of He-I operation. In fact, at low current, the current sharing and critical temperatures are well above the transition temperature $T_\lambda$ from He-II to He-I (around 2 K). Heating of the strands up to current sharing implies that the surrounding helium undergoes the He-II to He-I phase transition, and the stability margin is thus governed by heat transfer in He-I. At the ill-cooled transition, however, the stability margin shows a peculiar behaviour. Owing to the large heat transfer capability in He-II, the power balance at the strand surface remains favourable for recovery as long as the wetting helium is in the He-II phase. Therefore, in a first approximation, the full heat sink between the initial operating point and the transition temperature $T_\lambda$ is still available at levels of the operating current at which the conductor would have become ill-cooled for operation in He-I. In other words, the conductor can still be considered as *well-cooled* for temperature excursions up to $T_\lambda$. As the helium undergoes a phase transition at temperature $T_\lambda$, the available heat sink is significant, of the order of 200 mJ·cm$^{-3}$ of the helium volume. Finally, with increasing current, the power balance can again become unfavourable, as soon as the heat flux limits in He-II are reached. There, the final transition to the ill-cooled regime of operation takes place. This behaviour is shown in Fig. 40, following measurements by Lottin and Miller [41].

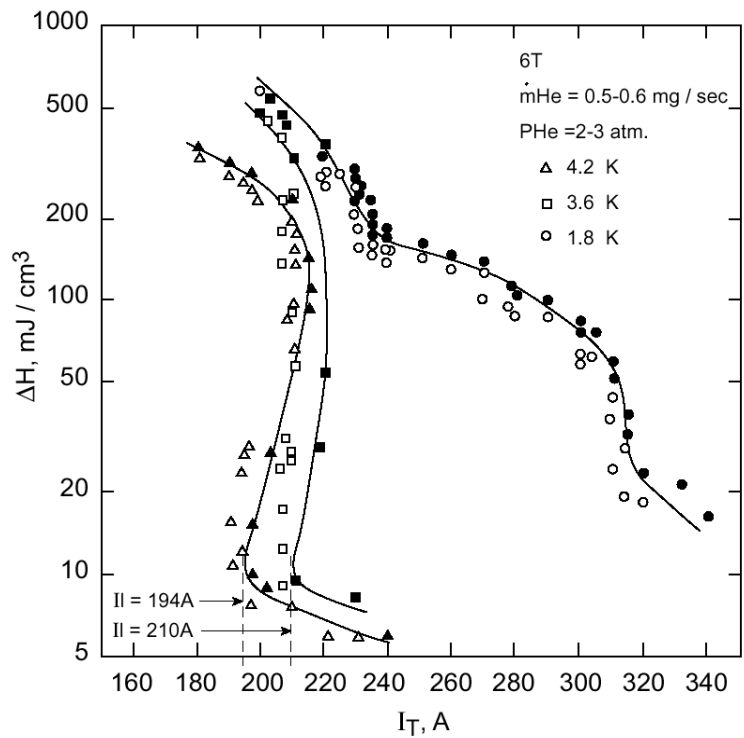

**Fig. 40:** The stability margin of a Nb–Ti CICC as a function of the operating current, measured by Lottin and Miller [41], at different temperatures in supercritical and superfluid helium. (Reproduced from [41] by permission of the IEEE. Copyright 1983 IEEE.)

## 6.5 Minimum propagating zones

The discussion so far has examined normal zones extending over large lengths of superconductor, ideally as large as the whole winding length. In reality, normal zones are established by small energy inputs over limited lengths of superconductor. In order to withstand these inputs, the cryostability and equal-area conditions would require a much too severe limitation on the operating current density. A better design criterion can be identified by resorting to the concept of the minimum propagating zone, originally defined by Wipf [43]. This concept was developed to a great extent by Wilson and Iwasa [44]. Following Wilson, we again consider the case of a superconducting wire with a normal zone in the centre. Using the auxiliary variable $S$ to represent the heat flux along the wire, because of symmetry at the centre of the normal zone we must have that:

$$S = 0, \qquad (53)$$

irrespective of the temperature reached by the superconductor. In this case we can again use the equal-area theorem and state that an equilibrium condition is defined by Eq. (46) (neglecting the variation of the thermal conductivity with temperature) where now, however, the central temperature $T'_{eq}$ is no longer $T_{eq}$ but, rather, is defined as the temperature at which the equal-area condition is satisfied. This situation is shown graphically in Fig. 41.

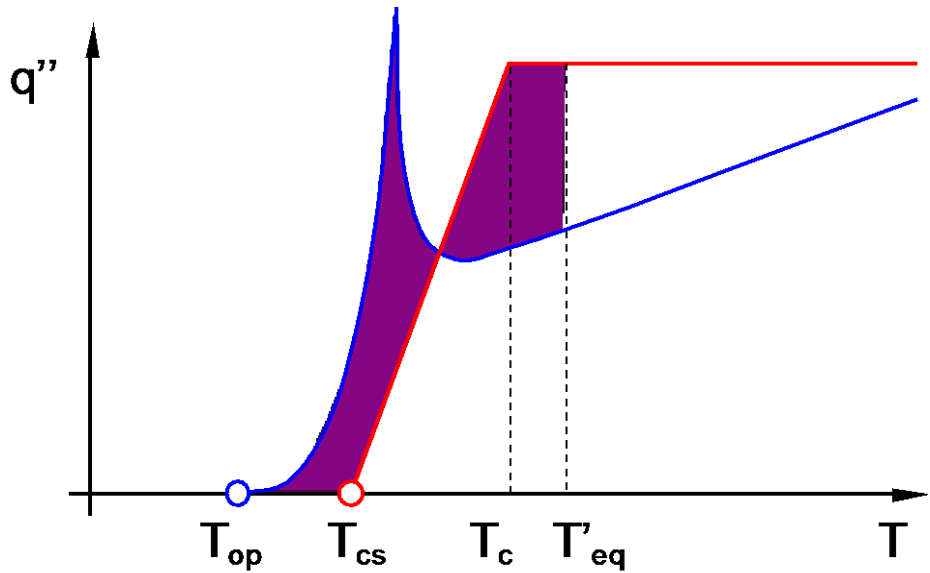

**Fig. 41:** A graphical interpretation of the equal-area theorem in the case of heat generation above the maximum steady state allowed by the equal-area theorem. The superconductor temperature profile with maximum temperature $T'_{eq}$ is an equilibrium point, as it also satisfies the equal-area condition, but it is unstable.

For a choice of heat generation above that allowed for steady-state equilibrium by the equal-area theorem, the central temperature $T'_{eq}$ must be lower than $T_{eq}$. The corresponding temperature profile in space can be found by integrating numerically Eq. (46). A family of such profiles, as produced by Wilson [44], is shown in Fig. 42. Each curve in the family corresponds to a different generation curve, and thus to a different equilibrium temperature. A property of the temperature profiles thus obtained is that for a given heat generation and cooling condition, any normal zone with a temperature profile below the one obtained will collapse because the cooling exceeds the generation, and the superconductor will recover from the local transition. If the temperature profile of a normal zone is above that obtained by the equal-area condition, then the normal zone will grow in time, leading to thermal runaway. Hence the normal zone identified by this modified equal-area condition represents an unstable equilibrium point, determining the boundary between recovery and thermal runaway. Because of this, it has been called the Minimum Propagating Zone (MPZ) [43].

The MPZ concept makes it possible to estimate the energy margin of the superconductor against short energy perturbations. It was observed by Wilson that if an energy input has a dimension in space smaller than the MPZ length, then the temperature profile evolves quickly towards the MPZ profile. This led him to postulate that the energy margin can be estimated as the energy necessary to instantaneously establish the MPZ. This can be regarded, in fact, as the minimum energy necessary in all conditions to quench the conductor, or the Minimum Quench Energy (MQE), and is therefore a conservative estimate for the energy margin. Any energy input happening on a finite time-scale will be associated with heat transport and will result in an energy margin larger than the MQE.

Heat conduction in more than one dimension, as used to establish the heat balance, has a similar effect, providing additional cooling for the MPZ. A demonstration of this is shown in Fig. 43, which reports the measured energy margin versus the results of calculations in 1D and 2D geometry. The agreement with the 2D calculation is evident. Also it is clear that, as expected, the energy margin is larger than the MQE as computed from the above theory.

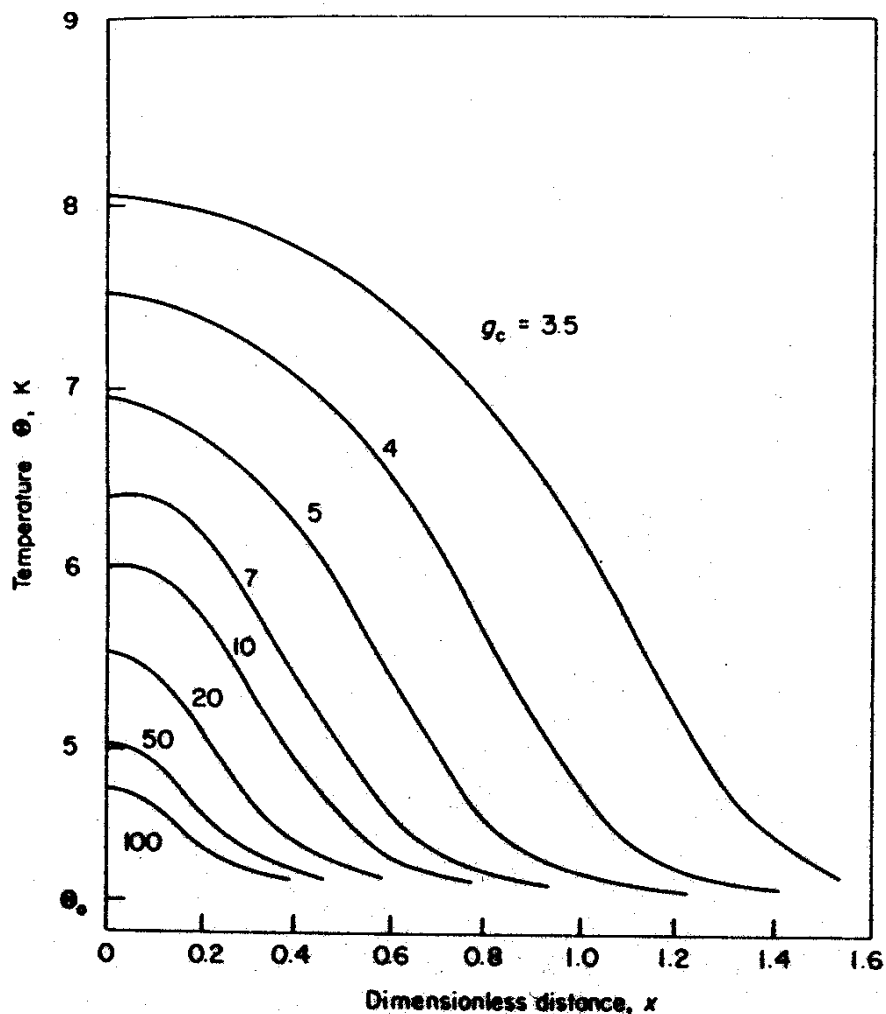

**Fig. 42:** The family of temperature profiles corresponding to MPZs obtained for different Joule heat generation conditions. Each curve represents the unstable equilibrium boundary between recovery and thermal runaway. (Reproduced from [44].)

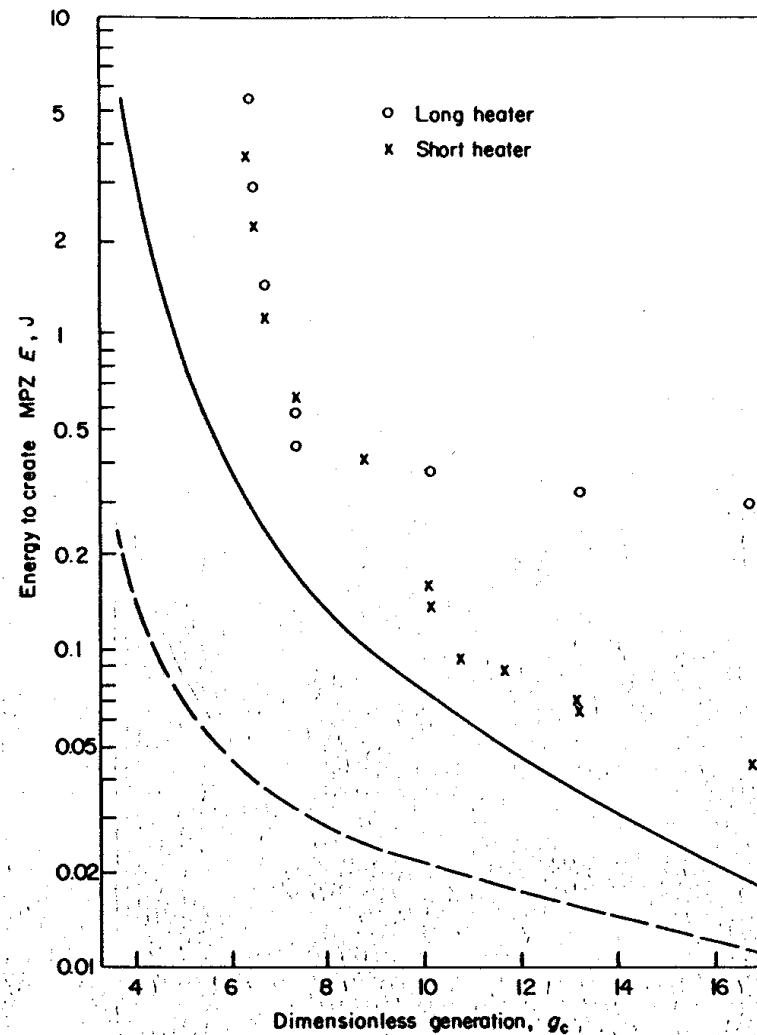

**Fig. 43:** The measured and computed energy margins for a small solenoid equipped with heaters, plotted as a function of the dimensionless Joule heat generation (after [44]). The computed curves refer to the MQE following either a 1D or a 2D calculation. (Reproduced from [44].)

### 6.6 Transient stability in the general case

At present, fully cryostable magnets are rarely the preferred designer choice. In an efficient magnet design, the cable operating current density must be kept high to make the magnet cross-section as small as possible. For a specified field or stored energy, and thus a given magnetomotive force, a maximum current density results in decreased material and production costs. As we have shown previously, a cryostable magnet needs a large amount of copper stabilizer – compared to the amount of superconductor – and a large amount of helium providing an ideally infinite heat sink. Therefore, a cryostable magnet has an intrinsically low operating current density.

On the other hand, cryostability implies that the conductor is stable against *any* disturbance spectrum, independent of the magnet details and the operating mode. In reality, the variety of conductor designs and of magnet winding techniques, together with the variety of operating requirements, results in a wide range of possible disturbance spectra. A cryostable conductor design is therefore, in general, excessively safe. Indeed, most magnets presently designed and built are not cryostable at the operating point, but they can still be operated reliably. The common feature of these magnets is that their stability margin is above the disturbance spectrum experienced during operation.

The first step in a sound design is thus to estimate the envelope of the perturbations that will be experienced. Subsequently, the conductor can be designed to accommodate these perturbations by means of a sufficiently large stability margin. Note that this process can imply iterations, as the disturbance spectrum can depend on the conductor and the coil design themselves.

Depending on the energy release dominating the disturbance spectrum, the different stabilization principles discussed in the previous sections can be used. A magnet operated in steady-state mode, with a tightly packed winding, affected by small mechanical disturbances localized in time and space (e.g. in the case of fully impregnated windings) may rely on the heat sink provided by the small enthalpy margin of the superconductor and stabilizer themselves: an *adiabatic winding*. To stabilize larger perturbations, the additional heat sink provided by helium may be necessary. Bringing helium into close contact with the conductor thus increases its stability margin, provided that the heat transfer at the wetted surface is efficient in the time-scale of the energy deposition considered. Magnets with small amounts of added helium (or other heat sinks) are called *quasi-adiabatic*, as they would in any case behave adiabatically for a fast enough time-scale. The stability margin can be made larger by increasing the heat sink (e.g. the amount of helium) and its efficiency in absorbing heat inputs (i.e. the heat transfer). This is typically the route followed in CICCs for large, pulsed magnets that are designed for use in energy storage or thermonuclear fusion applications. The disturbance spectrum is dominated in these cases by electromagnetic energy coupling through a.c. losses, which are generally much larger than the enthalpy margin of the superconducting wire itself. Several options are possible to increase the amount of helium and the heat transfer. In a forced-flow conductor, for instance, the helium flows in channels inside the conductor, and the strands are subdivided to increase their wetted perimeter and improve turbulent heat transfer. Another option is to use superfluid helium, which has an exceedingly high heat transfer rate, in close contact with the wire. In any case, the superconducting cable is in a *meta-stable* situation; namely, it can be quenched by a large enough energy input. The art consists in reaching the desired stability margin for reliable operation with maximum operating current density.

The main difficulty lies in the fact that the calculation of the energy margin associated with a perturbation of arbitrary distribution in space and time is a complex matter. All of the theories discussed so far have underlying approximations and limits, and the only way to attack the general case is by numerical simulation of the non-linear heat balance. Even so, the calculation remains a difficult task, involving accurate computation of heat conduction and possibly compressible helium flow in complex geometry, taking into account the non-linear material properties. In practice, the numerical calculation of the stability margin is the virtual analogue of an experiment, proceeding by trial and error to refine the approximation between the lower perturbation boundary, leading to recovery, and the upper boundary, resulting in a quench. The techniques discussed in the previous sections, involving verification of the power balance and of the enthalpy margin, provide approximate calculations that are usually sufficient for scoping calculations and design, and to start the search for more intense numerical calculations. The following examples give the typical logic sequence followed to achieve stable operation of magnetic systems in all operating conditions.

### 6.6.1  *The transient stability of the EU–LCT coil*

The Euratom–Large Coil Task (EU–LCT) coil was built in the framework of the Large Coil Task project, a multinational effort to demonstrate the feasibility of a toroidal field system for a thermonuclear fusion reactor [45]. The coil was wound in a D-shape, using the two-in-hand technique in seven double pancakes. The winding pack was epoxy impregnated under vacuum and enclosed in a thick steel casing, which provided the main support against the electromagnetic forces (see Fig. 44). At the nominal operating current of 11 400 A, the maximum field produced in the winding during full-array tests in the IFSMTF test facility was of 8.1 T, and the stored energy was about 100 MJ. The conductor itself, shown in Fig. 45, was obtained by *Roebel-cabling* 23 rectangular Nb–Ti strands (with copper stabilizer) around a central steel foil, and encasing this core in a steel jacket, producing a flat cable that was 10 mm thick and 40 mm wide. The helium could flow between each strand within the leak-tight jacket. Each strand, 2.35 × 3.1 mm² in size, contained 774 Nb–Ti filaments with a nominal diameter of 45 μm.

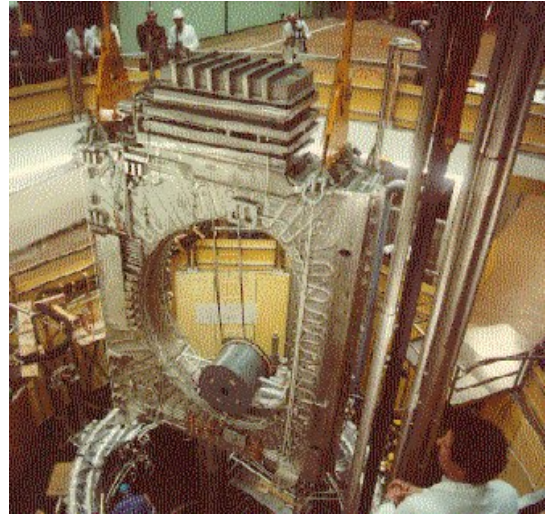

**Fig. 44:** The EU–LCT coil in its casing

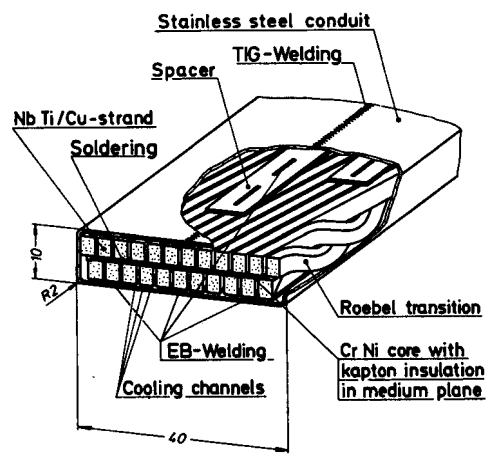

**Fig. 45:** The EU–LCT cable

The current density in the strands was around 70 A·mm$^{-2}$ in nominal conditions. This value is more than twice as high as the one for the BEBC conductor described earlier, and with an increase in the operating field from about 5 T in the BEBC magnet to about 8 T in the EU–LCT coil. The cooled perimeter of this complex configuration was estimated to be of the order of 165 mm, and at the nominal flow conditions the heat transfer coefficient was approximately 600 W·m$^{-2}$·K$^{-1}$. If we calculate the Stekly coefficient for these specific conditions, we obtain a value of $\alpha \approx 4$, considerably above the cryostable limit.

The disturbance spectrum during the operation of a TF coil in a fusion experiment is expected to be dominated by a.c. loss deposition during the field change associated with the sudden instability of the plasma column, or plasma *disruption*. Tests were performed on the EU–LCT coil, pulsing an external coil and producing field changes up to 0.3 T with a time-scale of 0.5 s. This deposited in the conductor energy of the order of 15 mJ·cm$^{-3}$ of strand, without causing a quench [45]. Calculations and measurements showed that for heat inputs in a short time-scale (0.5 ms), the stability margin was of the order of 10–30 mJ·cm$^{-3}$ of strand [46] in conditions comparable to the operating point of the cable. Over longer time-scales the stability margin increased, as more time was available to transfer heat to the helium.

No measurements are available for the conditions of the field pulse test quoted above, but a rough estimate, considering that the stability margin scales as the square root of the time-scale of the energy deposition [47], results in a minimum stability margin of the order of 100 mJ·cm$^{-3}$, well above the energy deposited by a.c. loss. Indeed, the coil never had a spontaneous quench during testing.

*6.6.2    The Tore Supra toroidal field magnet*

The Tore Supra [48] is a tokamak built in the 1880s at the Centre d'Etudes de Cadarache (France). Its toroidal field (TF) magnet is completely superconducting, and operates in a stagnant superfluid helium bath. The TF magnet is composed of 18 circular coils, wound out of a monolithic composite conductor in 26 double pancakes. The double pancakes are separated by spacers that maintain electrical insulation but allow the free flow of helium around the conductor and ensure a helium percentage in the winding pack of the order of 50% of the conductor volume. The winding pack is kept under compression by an external steel casing that provides the tightness for the superfluid helium bath, which is maintained at a temperature of approximately 1.8 K and a pressure of about 1 bar in normal operating conditions. At the operating current of 1400 A, the maximum field produced on the winding pack is of 9 T, for a stored energy in the TF magnet of 610 MJ. The conductor (see Fig. 46) is a rectangular wire, of dimensions 2.8 × 5.6 mm$^2$, with 11 000 Nb–Ti filaments of 23 μm diameter in a mixed copper and CuNi matrix. The nominal Nb–Ti cross-section is 4.6 mm$^2$ and the copper cross-section is 10 mm$^2$. At the operating conditions, the current density in the wire is approximately 90 A·mm$^{-2}$.

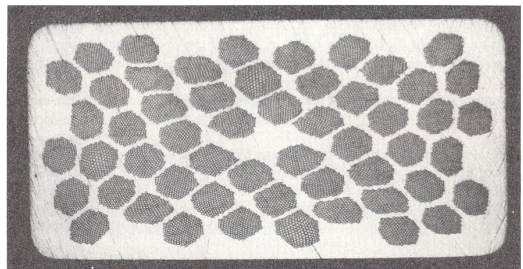

**Fig. 46:** The Tore Supra strand

For the Tore Supra, the tolerance against the disturbance spectrum was formulated with the requirement that that the conductor must be able to recover:

- after a localized (length of the order of some millimetres) temperature excursion up to 30 K, or
- after a global (one full pancake) temperature excursion to 15 K, or
- after a plasma current disruption when the conductor is subjected to a field change of 0.6 T in 10–20 ms.

Stability in superfluid helium has peculiar characteristics as compared to the situation of a conductor wetted by boiling or supercritical normal helium. The main difference is the large heat transfer capability of superfluid helium [49]. At small heat fluxes, the heat transport in superfluid helium is virtually infinite and the heat transfer coefficient $h$ from the conductor to the helium is mostly governed by the Kapitza resistance at the wetted surface, with rather large values, in the range of several thousands of W·m$^{-2}$·K$^{-1}$ [49]. The picture is different for large heat fluxes. In both steady-state and transient conditions, there is an upper limit to the heat flux that can be supported by superfluid helium before reaching the transition to the normal state, the so-called *lambda* line. This limit depends both on the helium state and on the geometry. In the case of 1 bar sub-cooled superfluid helium, the operating condition of the Tore Supra, a normal helium film forms at the wetted surface as soon as the peak heat flux is exceeded [50]. At the same time, the heat transfer drops while the conductor temperature rises sharply.

The consequence is that for small heat fluxes – for example, those deriving from mechanical energy releases – the heat removal is such that the helium heat capacity available for stabilization can be used completely. Larger energy depositions can be tolerated until the associated heat flux is below the maximum allowable value. This limits the available heat sink, as seen from the conductor side, to a fraction of the total helium volume. In the case of the Tore Supra, calculations and experiments were performed to guarantee that the conditions given above could be satisfied. In particular, a 60 m long cable had been tested in conditions comparable to the operation of the TF coil [51]. It was found that at the nominal operating current of 1400 A, the cable was stable against a field pulse (1 T in 8 ms) comparable to the one required in the design specifications. The a.c. loss deposited by this field pulse was around 35 mJ·cm$^{-3}$ of wire, and in these conditions no normal zone could be detected. The average heat flux associated with such an a.c. loss is approximately 5 kW·m$^{-2}$. This value, for the geometry of the cooling channel of the Tore Supra conductor, is well below the critical heat flux limit, which can be estimated to be of the order of 100 kW·m$^{-2}$ [49]. As the heat flux does not limit heat transfer, practically all the helium enthalpy from the operating temperature to the lambda transition is used for stabilization. The typical helium enthalpy from 1.8 K to $T_\lambda$ is of the order of 300 J·m$^{-3}$ of the helium volume, which is approximately 150 J·m$^{-3}$ of the strand volume. This last value is a good estimate of the stability margin in normal operating conditions.

# 7    Summary and advanced topics

This chapter has presented the basic considerations and models that go into the achievement of stable superconductors. Overall, we can see the strategies presented above as a trade-off between the desired performance and the allocated margin. One way to see this is to look at the schematic representation in Fig. 47, where the various stabilization strategies discussed are plotted in terms of the typical range of energy margins versus the typical range of operating current densities for which the strategy can be applied. The reader is warned that, as for the perturbation spectrum, this is only an order-of-magnitude representation, and exceptions can deviate considerably from the ranges identified there. Overall, however, we see that a high operating current density is invariably associated with a small energy margin. This implies that much effort must be put into the control and reduction of the perturbation spectrum.

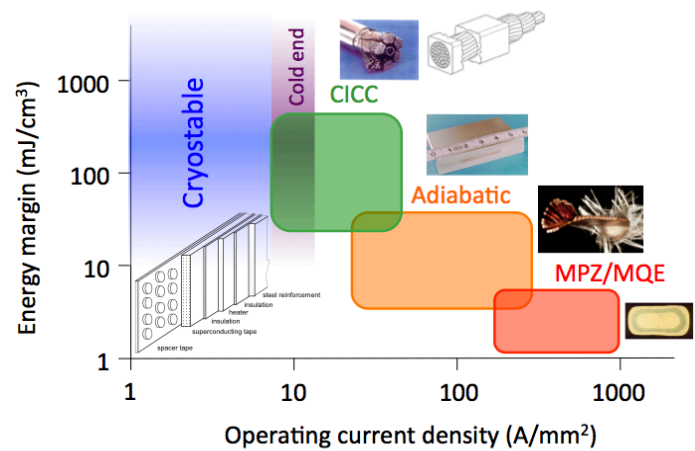

**Fig. 47:** A scatter plot of the typical range of energy margins versus operating current densities corresponding to the various stabilization strategies discussed in this chapter.

Enough is known about the mechanisms determining stability so that, in conjunction with other constraints, superconductors can be designed and optimized successfully. However, this does not mean that the field is not open to new areas of research. As new magnet designs are proposed, and as more stringent requirements are imposed on the designer, areas of further study continue to open up: in particular, work towards improving our understanding of stability under transient operating conditions, and the interaction of magnetic and thermal instabilities.

The details of transient local heat transfer are not fully known, nor understood, especially in complex flow geometries such as are often used for CICCs with cooling passages. As heat transfer plays such an important role in the determination of the stable behaviour of a superconductor, this point is somewhat surprising, but must be understood in terms of the difficulties inherent in the precise measurement of flow and heat transport in a cryogenic fluid.

Stability depends in a synergistic manner on the d.c. and a.c. operating conditions of the cable in the coil. This is a main direction of research in the field of stability. In particular, in view of the applications to pulsed magnets, the interaction of stability, current distribution and a.c. losses in the cable is one of the main topics. The so-called *ramp-rate* limit of operation for pulsed magnets (a decrease in the maximum achievable current at increasing field change rate) is an outstanding example of this synergistic interaction. The appearance of such a phenomenon, explained so far in terms of non-uniform current distribution and a degradation of the stability margin of the cable, has alerted us to the difference between d.c. stability, with a constant operating current and background field, and a.c. stability of the cable.

The distribution – and redistribution – of current among the strands and within the cable can have dramatic effects on stability. This statement applies to most cables used in technical applications (flat cables, CICCs and *super-stabilized* cables). Certainly, the general solution of the thermal, hydraulic and electromagnetic behaviour of a cable can be regarded as a formidable task. For this reason, most of the efforts to understand current distribution in multistrand cables have been limited so far to the purely electromagnetic problem, neglecting the intrinsic coupling with the thermal behaviour [52–54]. Only recently there have been more general attempts to consistently solve the coupled electromagnetic and thermal problem, and models have been presented for triplet of strands [55] and flat accelerator cables [56].

During a thermal transient, the current in a quenched strand tends to redistribute to the neighbouring strands, driven by the voltage of the normal zone. The redistribution takes place across the transverse contact resistance (or at the joints in the case of insulated strands). The variation in the strand current induces a change in the Joule heating rate, coupling back to the temperature evolution. To model the redistribution process, mutual inductive coupling of strands must be taken into account, while capacitive effects are negligible. Because a cable is strongly non-isotropic and because it has discrete contacts at the strand crossing, the first natural approach to a model of the current distribution is the use of an electrical network modelling the strands as uniform current density sticks, coupled inductively and through localized cross-resistances (see, e.g., [52, 54]). This *network* approach is solved by Kirchhoff's voltage and current laws, and requires that appropriate current *loops* are set for each degree of freedom in the cable cross-section. It is very detailed, providing information on each strand crossover contact, but it can result in a very large number of equations that are not conveniently coupled to a system of partial differential equations such as those given above.

One alternative, which has been used extensively for analytical studies, is to approximate the cross-contacts as a continuous transverse conductance (see, e.g., [53]). A typical example is that of an ideal two-strand cable. In this case, the governing equations become identical to those for an electrical transmission line with negligible capacitance, a well-known problem in electromagnetics. This semi-continuum approach is also useful for stability studies.

Super-stabilized superconductors, used in large magnets with low-intensity perturbation spectrum (detector magnets for high-energy physics, or SMES magnets), are a special field that is complex, but rather well understood. In super-stabilized conductors, a large amount of high-conductivity material is added in parallel to the cable for protection. The distance of the stabilizer from the multifilamentary area, and its low resistivity, result in an increase of the current diffusion time out of the superconductor into the stabilizer. This effect is negligible within a strand, but becomes appreciable in the limit of large segregated stabilizers, when this time can become comparable or larger than the time-scale of the evolution of the thermal transient. The cable is said to be *super-stabilized* if the time needed for current distribution is comparable to or larger than the time of flight of the normal zone along the same section of conductor. In this type of cable, the power dissipated by Joule heating during a transition to the normal state is initially much higher than the value reached after the current diffusion has taken place. After complete current diffusion, the heating decreases to the asymptotic steady-state value corresponding to a uniform current distribution. The variation of Joule heating associated with the current diffusion affects the recovery of the cable. Furthermore, the current diffusion can cause multiple stability boundaries, as well as stationary and travelling normal zones. Stability models for super-stabilized cables are obviously focused on the effect of current distribution inside the massive stabilizer. Continuum models are commonly used to describe this process [57, 58]. The details of the superconducting cable, as well as heat transfer to the helium, are lesser issues.

Finally, the old problem associated with flux jumps, dating back to the beginning of the history of superconducting magnet technology, should not be forgotten. Indeed, the push for higher fields in compact magnets such as accelerator dipoles and quadrupoles drives the need for current density to very high values in $Nb_3Sn$, in excess of 3000 A·mm$^{-2}$ at 12 T and 4.2 K. At the same time, for manufacturing reasons, the filaments of these high-$J_c$ materials are as large as 100 μm, and at present cannot be made much smaller than 30 μm. Such a high $J_c$ and a large diameter causes flux jumps at low field (0–2 T), as should be expected. An unexpected additional problem is associated with a newly identified *self-field instability*, which can appear at intermediate and high fields (4–12 T) in strands of large critical current and diameter, of the order of 1 mm or larger [59]. Such strands are required for large-scale applications of high-field magnets, to help increase the final size of the cable and ease protection. The drawback of a large strand diameter, associated with the very high values of $J_c$ quoted earlier, is that the transport current tends to remain confined in a very thin skin of filaments, at the periphery of the multifilamentary composite. A simple way to understand this instability is to consider this current distribution as generating a magnetic moment that can collapse and trigger an instability, in much the same way as a flux jump.